\documentclass[sigconf]{acmart}

\AtBeginDocument{%
  \providecommand\BibTeX{{%
    \normalfont B\kern-0.5em{\scshape i\kern-0.25em b}\kern-0.8em\TeX}}}

\setcopyright{acmcopyright}
\copyrightyear{2019}
\acmYear{2019}
\acmConference[SC '19]{The International Conference for High Performance Computing, Networking, Storage, and Analysis}{November 17--22, 2019}{Denver, CO, USA}
\acmBooktitle{The International Conference for High Performance Computing, Networking, Storage, and Analysis (SC '19), November 17--22, 2019, Denver, CO, USA}
\acmPrice{15.00}
\acmDOI{10.1145/3295500.3356163}
\acmISBN{978-1-4503-6229-0/19/11}

\usepackage{booktabs} 

\usepackage{amsmath,amssymb,amsfonts}
\usepackage{graphicx}
\usepackage{textcomp}
\usepackage{graphicx}
\usepackage{latexsym}
\usepackage{algorithm}
\usepackage{algpseudocode}
\usepackage{tabularx}
\usepackage{courier}
\usepackage{listings}
\usepackage{xcolor}
\usepackage{color,soul}
\usepackage{amssymb}
\usepackage{amsthm}
\usepackage{amsmath}
\usepackage{multicol}
\usepackage{lipsum}
\usepackage{comment}
\usepackage{amsmath}
\usepackage{subfig}
\usepackage[export]{adjustbox}
\usepackage{listings}
\usepackage[inline]{enumitem}
\usepackage{mathtools}
\usepackage{algpseudocode}
\usepackage{tabstackengine}
\usepackage{subfig}
\usepackage{listings}
\usepackage{multirow}
\usepackage{arydshln}
\usepackage{mathtools}
\usepackage{todonotes}
\usepackage{pifont}
\usepackage{algpseudocode}
\algrenewcommand\textproc{}
\usepackage[all]{nowidow}
\usepackage{array}
\usepackage{amssymb}
\usepackage{pifont}
\usepackage[skip=4pt,font=small]{caption}
\usepackage{amsmath}
\usepackage{bm}
\usepackage{tikz}
\usepackage{soul}

\newcommand{\tbhline}{\noalign{\hrule height 1pt}}
\newcommand\inner[2]{\langle #1, #2 \rangle}

\newcolumntype{?}{!{\vrule width 1pt}}
\newcolumntype{^}{!{\vrule width 1.2pt}}

\renewcommand{\O}[1]{$\mathcal{O}(#1)$}

\algtext*{EndWhile}
\algtext*{EndIf}

\newcommand{\xmark}{\ding{55}}%

\newcommand{\fourKperf}{30}
\newcommand{\eightKperf}{2}

\definecolor{myblue}{RGB}{91,155,213}
\definecolor{mydark}{RGB}{0,0,0}

\DeclareRobustCommand{\hllightgray}[1]{{\sethlcolor{lightgray}\hl{#1}}}

\begin{document}


\title[A Scalable High-resolution Image Reconstruction Framework]{iFDK: A Scalable Framework for Instant High-resolution Image Reconstruction}

\author{Peng Chen}
\orcid{1234-5678-9012}
\affiliation{%
  \institution{Tokyo Institute of Technology}
  \institution{AIST-Tokyo Tech Real World Big-Data Computation Open Innovation Laboratory, National Institute of Advanced Industrial Science and Technology}
}
\email{chen.p.aa@m.titech.ac.jp}

\author{Mohamed Wahib}
\orcid{1234-5678-9012}
\affiliation{%
  \institution{AIST-Tokyo Tech Real World Big-Data Computation Open Innovation Laboratory, National Institute of Advanced Industrial Science and Technology}
}
\email{mohamed.attia@aist.go.jp}

\author{Shinichiro Takizawa}
\orcid{1234-5678-9012}
\affiliation{%
  \institution{AIST-Tokyo Tech Real World Big-Data Computation Open Innovation Laboratory, National Institute of Advanced Industrial Science and Technology}
}
\email{shinichiro.takizawa@aist.go.jp}

\author{Ryousei Takano}
\orcid{1234-5678-9012}
\affiliation{%
  \institution{National Institute of Advanced Industrial Science and Technology}
}
\email{takano-ryousei@aist.go.jp}

\author{Satoshi Matsuoka}
\orcid{1234-5678-9012}
\affiliation{%
  \institution{Tokyo Institute of Technology}
  \institution{RIKEN Center for Computational Science, Hyogo, Japan}  
}
\email{matsu@acm.org}

\renewcommand{\shortauthors}{Chen, P. and Wahib, M., et al.}

\begin{abstract}
Computed Tomography (CT) is a widely used technology that requires compute-intense algorithms for image reconstruction. We propose a novel back-projection algorithm that reduces the projection computation cost to 1/6 of the standard algorithm. We also propose an efficient implementation that takes advantage of the heterogeneity of GPU-accelerated systems by overlapping the filtering and back-projection stages on CPUs and GPUs, respectively. Finally, we propose a distributed framework for high-resolution image reconstruction on state-of-the-art GPU-accelerated supercomputers. The framework relies on an elaborate interleave of MPI collective communication steps to achieve scalable communication. Evaluation on a single Tesla V100 GPU demonstrates that our back-projection kernel performs up to 1.6$\times$ faster than the standard FDK implementation. We also demonstrate the scalability and instantaneous CT capability of the distributed framework by using up to 2,048 V100 GPUs to solve a 4K and 8K problems within \fourKperf~seconds and \eightKperf~minutes, respectively (including I/O). 
\end{abstract}

%
%



\maketitle

\section{Introduction} \label{sec:Introduction}

High-resolution Compute Tomography (CT) is a technology used in a wide variety of fields, e.g. medical diagnosis, non-invasive inspection~\cite{shimadzu:4k}, and reverse engineering~\cite{ProScan:ReverseEngineering, GOM-CT}. In the past decades, the size of a single three-dimensional (3D) volume generated by CT systems has increased from hundreds of megabytes (the typical sizes of a volume are $256^3$, $512^3$) to several gigabytes (i.e. $2048^3$, $4096^3$)~\cite{thompson2011rapid,myers2016high, betzefficient}. The increased demand for
rapid tomography reconstruction 
and the associated high computational cost attracted heavy attention and efforts from the HPC community~\cite{yang2006parallel,thompson2011rapid, gregor2011distributed, cui2013distributed, rosen2013iterative, palenstijn2015distributed, Bicer:Rapid-Tomographic,wang2017massively, sabne2017model, 8425161, Blas:2013:PIX:2488551.2488589}.
{
As illustrated in~\cite{pan2009commercial}, the FDK~\footnote{Feldkamp, Davis, and Kress~\cite{feldkamp1984practical} presented a convolution-backprojection formulation (known as FDK algorithm) for CT image reconstruction in 1984. 
FDK is also known as the Filtered Back Projection (FBP) algorithm.} algorithm is widely regarded as the primary method to reconstruct 3D images (or volumes) from projections, i.e. X-ray images.
}
The FDK algorithm includes a filtering stage (also known as convolution) and a back-projection stage. The computational complexities of those two stages are \O{N^2log(N)} and \O{N^4}, respectively.
Researchers are increasingly relying on the latest accelerators to improve the computational performance of FDK, e.g. Application Specific Integrated Circuits (ASIC)~\cite{wu1991asic}, Field-Programming Gate Array (FPGA)~\cite{coric2002parallel,xue2006acceleration,subramanian2009c,henry2012fpga}, Digital Signal Processor (DSP)~\cite{liang2010optimized}, Intel Xeon-Phi~\cite{rohkohl2009rabbitct}, Multi-core CPUs~\cite{wang2017massively}, and Graphics Processing Unit (GPU)~\cite{xu2005accelerating,rezvani2007ff,zhao2009gpu,zinsser2013systematic}.
This paper focuses on GPU-accelerated supercomputers for two reasons. First, GPUs are dominantly used for tomographic image reconstruction~\cite{sabne2017model,8425161,despres2017review,sharp2007gpu,xu2007real,jia2011gpu}. Second, GPU-accelerated supercomputers are increasingly gaining ground in top-tier HPC systems.

Instantaneous high-resolution image reconstruction, i.e. generating a volume moments after processing the scanned image projections, has long been the holy grail of CT technologies. The following are some of the challenges one has to consider when targeting instant high-resolution image reconstruction.
First, FDK is a well-researched algorithm, yet it remains a fact that FDK, with its high-compute intensity, can benefit from innovation at the algorithm level in order to reduce the cost of computing the projections (while preserving computational precision). Second, CPUs and GPUs have different architectures. A fully heterogeneous solution necessitates careful assignment and orchestration of computing tasks to CPUs and GPUs.
Third, high-resolution image reconstruction is limited by GPU memory capacity.
{
Taking a volume of size $4096^3$ for instance, the required storage is 256GB, which largely exceeds the memory capacity of a single GPU.} Hence a distributed implementation is essential to avoid the GPU memory capacity bottleneck. 
Fourth, the effective use of MPI over the system hierarchy is critical to optimize data movement within the system. 
Finally, non-trivial optimizations are required to design an end-to-end pipeline for the computation of the filtering and back-projection stages using thousands of CPUs/GPUs (and this includes the I/O bottleneck of the parallel file system).

We target to enable instantaneous high-resolution image reconstruction while also providing the technical capacity required for advancing to unprecedented resolutions, e.g. $8192^3$. We propose a scalable framework, called iFDK, for computing FDK on GPU-accelerated supercomputers.
Optimizing the back-projection stage is crucial since back-projection is the computational bottleneck in most of the practical CT image reconstruction algorithms. 
We propose a novel back-projection algorithm that reduces the number of operations for computing the projection computations to a factor of 1/6 from the standard FDK algorithm. The algorithm is also general and thus can be adopted by iterative reconstruction methods, in which the back-projection is required to be repeated dozens of times, e.g. ART~\cite{gordon1970algebraic}, SART~\cite{andersen1984simultaneous}, MLEM~\cite{shepp1982maximum}, and MBIR~\cite{7105892}. 

Prevalent approaches in the literature execute both the filtering and back-projection stages on GPUs. Contrarily, we improve the efficiency by taking full advantage of the heterogeneity in GPU-accelerated supercomputers. The filtering stage is executed on CPUs with optimizations for multi-threading and SIMD vectorization~\cite{cockshott2013simd}. The back-projection stage is executed on GPUs with optimizations at the algorithmic level to reduce the cost of the projection computations while also improving the data locality. 


The proposed framework is further optimized for reducing communication. More specifically, we propose a scalable problem decomposition scheme at which several independent sub-tasks are decomposed to a two-dimensional mesh of MPI ranks. The image reconstruction problem is decomposed such that the horizontal MPI ranks handle the input while the vertical ranks generate the output volume. Most importantly, the sub-tasks are overlapped in a pipelined fashion to be computed in parallel.


Using more than 2,000 Nvidia Tesla V100 GPUs, we solve the 4K ({$2048{\times}2048{\times}4096{\rightarrow}4096^3$}) and 8K ({$2048{\times}2048{\times}4096{\rightarrow}8192^3$}) high-resolution image reconstruction problems~\footnote{
The image reconstruction problem is defined in Section~\ref{terminology}.
} within {\fourKperf}~seconds and {\eightKperf}~minutes, respectively. This includes the end-to-end processing time: loading projections from the Parallel File System (PFS), the filtering stage, the back-projection stage, MPI communication, and finally storing the output 3D volume to the PFS. 

The contributions in this paper are as follows:
{
    \setlength{\leftmargini}{10 pt}
    \begin{itemize}
        \item We propose a novel back-projection algorithm that reduces the cost to compute the projections and improves cache locality. 
        \item We propose a scalable and distributed framework for high-resolution image reconstruction on heterogeneous supercomputers.
        \item We demonstrate that high-resolution image reconstruction problems can be solved within tens of seconds by iFDK. To the author's knowledge, this is the first attempt to achieve instant distributed CT image reconstruction for 4K and 8K resolutions.   
    \end{itemize}
}

The rest of this paper is organized as follows. 
In Section~\ref{sec:bkground}, we review the background of CUDA and FDK algorithm. 
In Section~\ref{proposed-fdk}, we propose the novel FDK algorithm and CUDA implementation.
In Section~\ref{sec:mnFDK}, we propose the distributed framework (namely iFDK) and present the performance model. 
Section~\ref{sec:Evaluation} describes the evaluation results. 
Section~\ref{sec:disc} discusses potential impact of iFDK in real-world.
In Section~\ref{sec:related-works}, we introduce the related work.
Finally, Section~\ref{sec:Conclusion} concludes.

\section{Background} \label{sec:bkground}
In this section, we introduce the basics of CUDA and describe the details of the FDK algorithm.
\subsection{CUDA}\label{sec:cuda}
Nvidia Compute Unified Device Architecture (CUDA) is a parallel computing platform and application programming model.
We briefly introduce the concepts of CUDA's architecture and memory hierarchy,
more details can be found in~\cite{cudaToolkit}.

{\bf{CUDA architecture.}}
CUDA is built on an array of multi-threaded Streaming Processors (SMs). Massive thread-level parallelism is abstracted into a hierarchy of threads running in a single instruction multi-thread (SIMT) fashion. The threads in CUDA are grouped into warps
(32 threads execute as a warp),
blocks, and grids. Thousands of threads are created, scheduled, and executed concurrently.

{\bf{CUDA memory hierarchy.}}
To approach the peak performance of GPU, it is essential to implement applications that efficiently utilize CUDA's memory hierarchy.
{
\begin{enumerate*}[label=(\Roman*)]
\item {\bf{Global memory}}. The largest off-chip memory. A coalesced access pattern is required for the CUDA kernels to achieve the highest bandwidth. 
\item {\bf{Shared memory}}. A fast on-chip scratchpad memory, which shares space with the L1 cache. The access scope is limited to a single CUDA block. 
\item\label{sec:const-mem} {\bf{Constant memory}}. A read-only constant cache shared by all of the SMs.
\item\label{sec:texture} {\bf{Texture memory}}. A read-only on-chip memory which is optimized for the spatial locality. The operation of reading a texture is also called texture fetch. Texture fetch~\cite{schwarz2011texture} can support efficient sub-pixel interpolation (as will be shown in Alg~\ref{alg:subpixel}). 
\item\label{sec:register} {\bf{Register files}}. The fastest on-chip and thread-private memory.
\end{enumerate*}
}

{\bf{CUDA shuffle intrinsic.}}
CUDA provides efficient intra-warp communication instructions, namely, shuffle. Without using shared or global memory, threads in a single warp can exchange registers directly. In terms of computing efficiency, using shuffle for in-register computation is superior in performance to the other memory types, due to its low latency and high throughput~\cite{chen2018efficient}. 
\begin{figure}[t]
  \begin{center}
    \includegraphics[clip,width=0.47\textwidth]{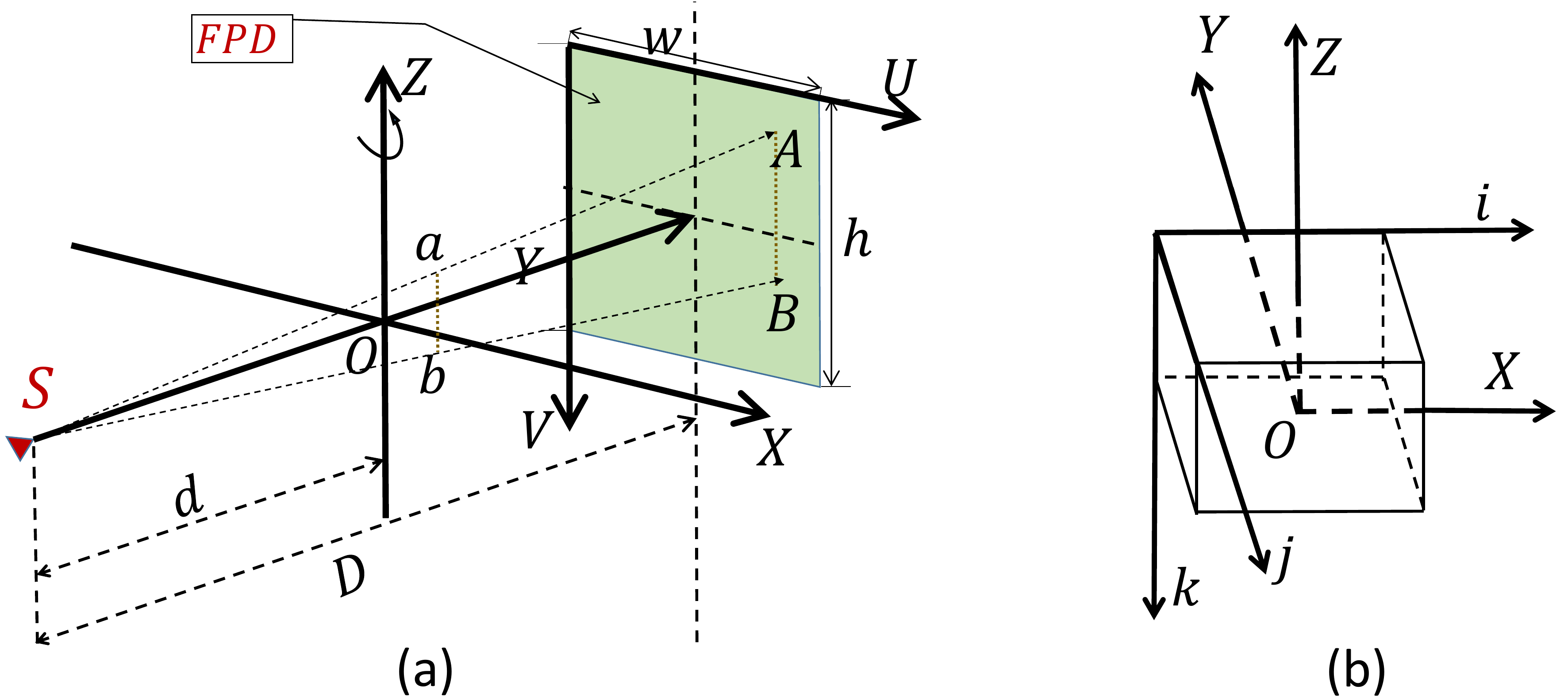}
    \caption{CBCT geometry and trajectory.}
    \label{fig:CT.geometry}
  \end{center}
\end{figure}
 \begin{table}[t]  
      \caption{CBCT parameter list.}        
      \scriptsize
      \centering
      \begin{tabular}{ ?c|l|c? }
      \tbhline
      \bf{Param} & \bf{Description} & \bf{Unit}\\
      \hline
      $N_p$ & the number of 2D projections & N/A  \\
      $N_u, N_v$ & the width and height of a 2D projection, respectively & pixel \\
      $D_u, D_v$ & FPD pixel pitch in U and V direction, respectively& mm/pixel\\
      $E_i, Q_i$ & the $i^{th}$ projections and the filtered result, respectively & N/A \\
      $F_{ramp}$ & 1-D Ramp filter~\cite{feldkamp1984practical} & N/A \\
      $F_{cos}$ & 2-D cosine table of size ($N_v, N_u$)~\cite{feldkamp1984practical} & N/A \\
      $P_i$ & the $i^{th}$ projection matrix of size 3$\times$4~\cite{rohkohl2009rabbitct,rit2014reconstruction} & N/A \\
      $d$ &  distance of X-ray source to rotation Z-axis&pixel\\
      $D$ & distance of X-ray source to FPD center & pixel\\
      $N_x, N_y, N_z$ & the number of voxels in X, Y, Z dimension, respectively & voxel\\
      $D_x, D_y, D_z$ & the pitch of volume  in X, Y, Z dimension, respectively & mm/voxel\\
      $\theta$ & rotation step angle, $\theta=2\cdot\pi/N_p$ & Rad\\
      $I$ & 3D volume & N/A \\
      \tbhline
       \end{tabular}  
       \label{tbl:cbct-param} 
\end{table}     

\subsection{FDK algorithm}\label{sec:fdk}
{
In this section, we revisit the 3D image reconstruction method (namely FDK) for Cone-Beam Computed Tomography (CBCT) as introduced by Feldkamp et al~\cite{feldkamp1984practical}. We briefly introduce geometry, arithmetic computation, and convolution. More details can be found in~\cite{avinash1988principles}.
}

\subsubsection{\bf{CBCT Geometry.}}\label{sec:geometry}
Figure~\ref{fig:CT.geometry} illustrates the CBCT geometry in detail and Table~\ref{tbl:cbct-param} lists all of the related parameters.
S is a micro-focus X-ray source, FPD (Flat Panel Detector) is a class of x-ray digital radiography detectors, which is principally similar to the image sensors used in digital photography. In addition, both the X-ray source and FPD are fixed relatively in position as Figure~\ref{fig:CT.geometry}a shows. Both of them rotate around the Z-axis while scanning objects placed at the center O to express a 3D volume as shown in Figure~\ref{fig:CT.geometry}b.

\subsubsection{\bf{FDK Computation.}}\label{sec:fdk-alg}
FDK is widely employed to build tomographic images in clinical and medical practice~\cite{pan2009commercial}.
FDK comprises a filtering stage (or convolution stage) as Algorithm~\ref{alg:filter} shows, and a back-projection stage shown in Algorithm~\ref{alg:bp}. 
In Algorithm~\ref{alg:filter}, cosine weighting ($F_{cos}$) and ramp filter ($F_{ramp}$) are convolved with projections ($E_{i}$) to generate the filtered result ($Q_i$). The details of $F_{cos}$ and $F_{ramp}$ (including improved versions) can be found in~
\cite{feldkamp1984practical,avinash1988principles}.
The shape of the $F_{ramp}$ filter deeply affects the final image quality, yet it has no effect on the compute intensity of the filtering stage. 

Algorithm~\ref{alg:bp} shows the back-projection algorithm. The projection matrix $P_i$, a well-aligned $3{\times}4$ matrix, incorporates all of the geometry information for the back-projection, i.e. \emph{d}, \emph{D}, $\theta$. More details on computing the  \emph{P} matrix are presented in literature~
\cite{wiesent2000enhanced,rit2014reconstruction}. 
It is important to mention the embarrassingly parallel nature of the back-projection computation when utilizing the projection matrix, in comparison to other methods in literature~\cite{serrano2014high,lu2016cache}.
As Algorithm~\ref{alg:subpixel} shows, the \emph{bilinear interpolation} method is adopted by most of the FDK implementations to fetch the intensity value of a 2D matrix with sub-pixel precision for updating each value of \emph{I} in Algorithm~\ref{alg:bp}~\cite{rit2014reconstruction,rohkohl2009rabbitct}.

\begin{algorithm}[t]
  \scriptsize 
  \caption{Filtering stage~\cite{avinash1988principles}.}
  \label{alg:filter}
  \begin{algorithmic}[1]
  	\Require{$E,F_{cos},F_{ramp},N_p,N_v,N_u$}
    \Ensure{$Q$}
    \For {$i{\in}[0,N_p)$}
	\State{$\tilde{E_i} \gets E_i \cdot F_{cos}$}\Comment{$\cdot$ means point-wise multiplication}
	\For {$j{\in}[0,h)$}
    	\State{$Q_i(j,\dots) \gets \tilde{E_i}(j,\dots){\otimes}F_{ramp} $}\Comment{${\otimes}$ means convolution}
    \EndFor
    \EndFor
  \end{algorithmic}
\end{algorithm}
  \begin{algorithm}[t]
    \scriptsize 
    \caption{Back-projection stage. This scheme is implemented in RTK~\cite{rit2014reconstruction}, RabbitCT ~\cite{rohkohl2009rabbitct}, etc.}
    \label{alg:bp}
    \begin{algorithmic}[1]
      \Require{$P, Q, N_p, N_x, N_y, N_z$}
      \Ensure{$I$}\Comment{generated 3D volume}
      \State{$I \gets 0$}\Comment{$I$ initialization}
      \For {$s{\in}[0,N_p)$}
          \For {$k{\in}[0,N_z)$}
              \For {$j{\in}[0,N_y)$}
                  \For {$i{\in}[0,N_x)$}
                      \State{\hllightgray{$[x,y,z]^T\gets P_s \cdot [i, j, k, 1]^T$}}\Comment{\textcolor{red}{3 inner product}}
                      \State{$f\gets 1/z$}
                      \State{$W_{dis}\gets f^2$}\Comment{distance weight}                    
                      \State{$[u, v]^T\gets [x,y]^T\cdot f$}\Comment{coordinates in FPD}                  
                      \State{$I(i,j,k) \gets I(i,j,k)+W_{dis}{\cdot}interp2(Q_s,u,v)$}
                  \EndFor
              \EndFor
          \EndFor
      \EndFor
    \end{algorithmic}
  \end{algorithm}
\begin{algorithm}[t]
  \caption{Bilinear interpolation with sub-pixel precision~\cite{jain1995machine}.}
  \scriptsize
  \label{alg:subpixel}
  \begin{algorithmic}[1]
    \Function{interp2}{$X,u,v$}\Comment{$X$ is 2D matrix, ($u$, $v$) is sub-pixel coordinate}
        \State{$[n_u,n_v]^T{\gets}[\mathrm{int}(u),\mathrm{int}(v)]^T$}\Comment{$n_u,n_v$ are integers}
        \State{$[d_u,d_v]^T{\gets}[u-n_u, v-n_v]^T$}\Comment{distance to left points}
        \State{$t_1{\gets}X(n_u, n_v){\cdot}(1-d_u)+X(n_u+1, n_v){\cdot}d_u$}\Comment{a sub-pixel value}
        \State{$t_2{\gets}X(n_u, n_v+1){\cdot}(1-d_u)+X(n_u+1, n_v+1){\cdot}d_u$}\Comment{a sub-pixel value}
        \State{\Return $t_1{\cdot}(1-d_v)+t_2{\cdot}d_v$}\Comment{final sub-pixel value}
    \EndFunction
  \end{algorithmic}
\end{algorithm}
\subsubsection{\bf{Convolution via FFT}}\label{sec:fft}
FFT~\cite{brigham1988fast} is essential to the filtering stage since one-dimensional FFT is used to perform the convolution operation in Algorithm~\ref{alg:filter}. 
For large problem sizes, FFT is typically the choice for the convolution computation~\cite{bracewell1995two}.
Optimized FFT primitives are typically provided by CPU/GPU vendors, e.g. Intel IPP (Intel Integrated Performance Primitives)~\cite{taylor2007optimizing}, 
and Nvidia cuFFT library~\cite{cudaToolkit}.
{
In Algorithm~\ref{alg:filter} line 4, the FFT primitive is employed to perform the convolution as follows. Regarding the two one-dimensional arrays, according to the Convolution Theorem~\cite{arfken1999mathematical}, convolution computation in the time domain equals point-wise dot product in the frequency domain. The reader can refer to a detailed illustration in~\cite{ConvolutionTheorem}.}


\subsection{Terminology}\label{terminology}
Throughout this paper, the image reconstruction problem and performance metric GUPS are defined as follows:
{
\vspace{-3pt}
\setlength{\leftmargini}{15 pt}
\begin{enumerate}[label=(\Roman*)]
\item\label{sec:def-problem} {\bf{$N_u{\times}N_v{\times}N_p{\rightarrow}N_x{\times}N_y{\times}N_z$}} is defined as the image reconstruction problem, where $N_u{\times}N_v{\times}N_p$ denotes the size of projections ({\bf{Input}}) and $N_x{\times}N_y{\times}N_z$ is the size of volume ({\bf{Output}}). 
\item\label{sec:def-GUPS} {\bf{GUPS}} is defined as giga-updates per second and is used as a performance metric. It is computed as {\small${\mathrm{GUPS}=\frac{N_x*N_y*N_z*N_p}{T*2^{30}}}$}, where $\small{\mathrm{T}}$ is the execution time (in seconds). 
\end{enumerate}
}

\section{Proposed Novel FDK algorithm}\label{proposed-fdk}
The next section discusses the execution of the filtering stage on the CPUs and the following section discusses the execution of the novel FDK algorithm (back-projection stage) on the GPUs. 
\subsection{Filtering Stage}\label{sec:filtering-algorithm}

\begin{figure}[t]
  \begin{center}
    \includegraphics[clip,width=0.485\textwidth]{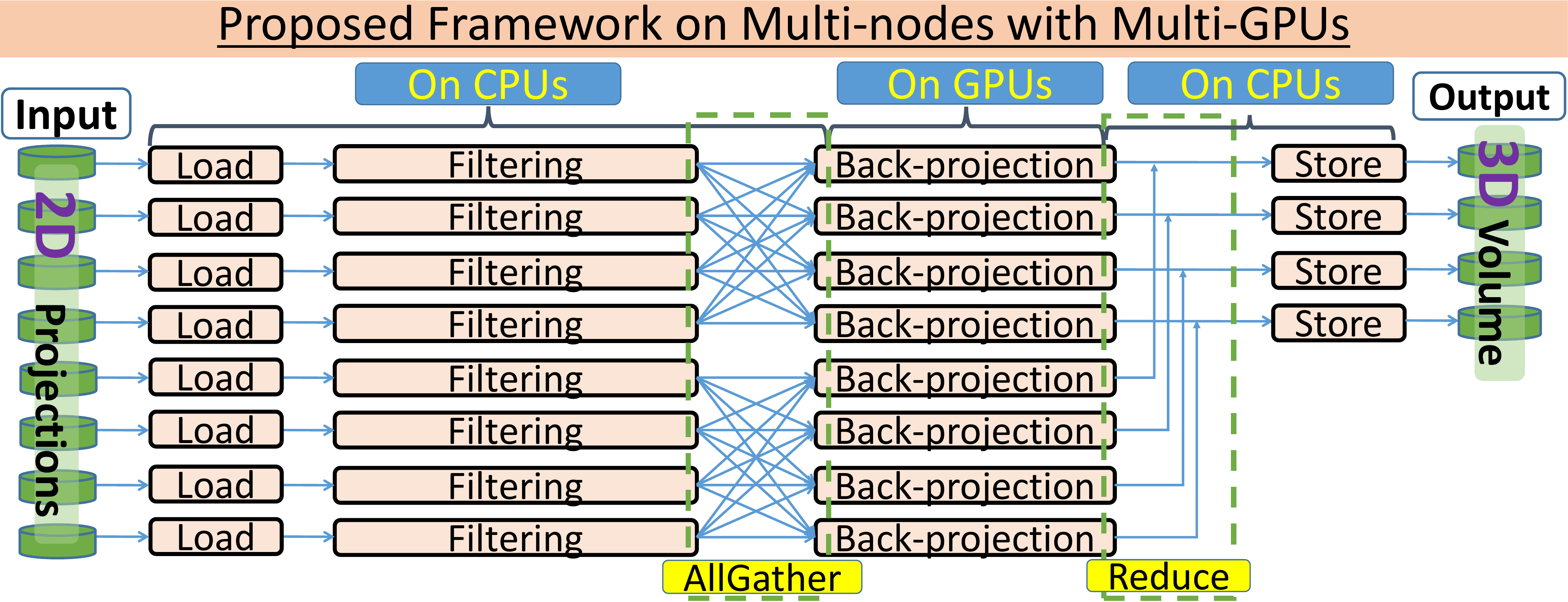}
    \caption{Overview of the proposed framework. Multi-nodes with GPUs accelerators are used such that a 3D volume is generated from 2D projections.
    }
    \label{fig:heterogeneous-implementation}
  \end{center}
\end{figure}

Figure~\ref{fig:heterogeneous-implementation} shows our heterogeneous computational flow, which is different from the typical method of using only the GPU for all the computation~\cite{okitsu2010high,blas2014surfing, lu2016cache}.
Utilizing the CPU to perform the filtering stage can be more efficient in comparison to using the GPUs to compute the entire FDK pipeline. {\bf{We list the three considerations}} that makes the fully heterogeneous model more efficient:
{
\begin{enumerate*}[label=(\Roman*)]
\item{\bf{Dedication of GPU to the compute intense stage: }} the latency of the filtering stage on the CPU can be hidden by overlapping it with the back-projection offloaded to the GPU. Hence, the GPU can be fully dedicated to the back-projection kernel (see Section~\ref{sec:cuda-bp}).

\item{\bf{Reduce memory pressure on GPU device memory:}} in the case that the GPU would be used for filtering, a batch of images have to be filtered at a time in order to fully utilize the GPU. The additional memory required to store the filtered projects would take away from the already limited GPU memory and would force the use of more GPUs in high-resolution problems to have enough aggregate memory capacity.

\item{\bf{Efficient communication:}} in our communication scheme, an AllGather collective is required after the filtering stage to fetch the filtered projections. When the filtering stage is applied at the GPU, the AllGather collective will be applied on data residing in GPU memory and not on data residing in the CPU memory (as is the case when the filtering is applied on the CPU). Applying AllGather on data residing on the GPU incur the extra cost of moving data across the PCIe interconnect, even when the GPUDirect~\cite{potluri2013efficient,li2019evaluating} Technology is enabled.
\end{enumerate*}
}
\subsection{Back-projection Stage} \label{sec:proposed-method}
We propose a general back-projection algorithm that reduces the computation and improves the data locality (regardless of the target architecture). 

\subsubsection{\bf{Theorems for Back-projection Algorithm}}
This section introduces three theorems which we use to propose a novel version of the original FDK algorithm at Algorithm~\ref{alg:bp}.
{
    \vspace{-3pt}
    \setlength{\leftmargini}{8 pt}
    \begin{itemize}
    \item {\bf{Theorem-1:}} As Figure~\ref{fig:CT.geometry} shows, when two 3D points \text{a} ($\tilde{i}$, $\tilde{j}$, $\tilde{k}$) and b ($\tilde{i}$, $\tilde{j}$, $\tilde{N_z}-k-1$) are symmetrical to XY plane, then the corresponding projection points A ($\tilde{u_A}$, $\tilde{v_A}$) and B ($\tilde{u_B}$, $\tilde{v_B}$) at FPD are symmetrical to the horizontal center line, namely $\tilde{u_A}=\tilde{u_B}$ and $\tilde{v_A}+\tilde{v_B}=N_v-1$. (proven by~\cite{zhao2009gpu}) 
    \item {\bf{Theorem-2:}} For points in the vertical line $\overline{ab}$ (parallel to Z-axis), their projection line $\overline{AB}$ is parallel to V-axis in FPD plane, namely the $\tilde{u}$ of line $\overline{AB}$ has a constant value. (the mathematical proof is trivial, we do not include it due to space constraints)
    \item {\bf{Theorem-3:}} When points are in the vertical line $\overline{ab}$ (parallel to Z-axis), such that computing the projection points use Equation~\ref{equ:uv}, then their \emph{z} is a constant value equal to $d+y_{ab}$, where $y_{ab}$ is the Y coordinate of line $\overline{ab}$. 
    (proof follows).
    \end{itemize}
}

{\bf{Proof of Theorem-3.}}\label{sec:proof-3}
We prove that in the specified rotation angle $\beta$ (or $i{\cdot}\theta$), if the $\hat{i}$ and $\hat{j}$ are fixed, the $z$ in Equ~\ref{equ:uv} is a constant value.
As Fig~\ref{fig:CT.geometry} shows, given a point \emph{a} of coordinate $(\tilde{i}, \tilde{j}, \tilde{k})$ in the volume coordinate system, its projection point A of coordinate (u, v) on the FPD can be computed using the projection equation as
\begin{equation}\small\begin{split}
	\label{equ:uv}
    \left\{
    \begin{alignedat}{4}
        [x,y,z]^T &= P_i\cdot[\tilde{i}, \tilde{j}, \tilde{k}, 1]^T\\
        [u,v]^T &= [x,y]^T\cdot 1/z\\
    \end{alignedat}
    \right.
\end{split}\end{equation}
where x,y,z are temporary variables. $P_i$ is a $3{\times}4$ projection matrix on the condition that the gantry rotation angle is $\beta$. Hence, $P_i$ may be written as
\begin{equation}\small\begin{split}
    \left\{
    \begin{alignedat}{4}
        \hat{P}_{i} &= M_1{\cdot}M_{rot}{\cdot}M_0\\
                    P_{i} &= \hat{P}_i[0:3]\\
    \end{alignedat}
    \right.
\end{split}\end{equation}
where the shapes of $\hat{P}_{i}$ and $P_{i}$ are $4\times4$ and $4\times3$, respectively. $M_0$, $M_{rot}$, and $M_1$ are listed as follows
\begin{equation*}
\setlength{\arraycolsep}{3pt}\small
\begin{split}
M_0=
\begin{pmatrix}
		D_x&      0&       0&      0\\
		0&       D_y&      0&      0\\
		0&       0&       D_z&     0\\
		0&       0&       0&       1\\
\end{pmatrix}
\cdot
\begin{pmatrix}
		1&  0&  0& -{(N_x-1)}/{2}\\
		0& -1&  0& {(N_y-1)}/{2}\\
		0&  0& -1& {(N_z-1)}/{2}\\
		0&  0&  0&            1 \\
\end{pmatrix}
\end{split}\end{equation*}

\begin{equation*}
\setlength{\arraycolsep}{3pt}\small
\begin{split}
M_{rot}=
\begin{pmatrix}
		1&       0&       0&       0\\
		0&       0&      -1&       0\\
		0&       1&       0&       d\\
		0&       0&       0&       1\\
\end{pmatrix}
\cdot
\begin{pmatrix}
		cos(\beta)&  -sin(\beta)&       0&       0\\
		sin(\beta)&   cos(\beta)&       0&       0\\
		0&                     0&       1&      0\\
		0&                     0&       0&       1
\end{pmatrix}
\end{split}\end{equation*}

\begin{equation*}
\setlength{\arraycolsep}{3pt}\small
\begin{split}
M_{1}=
\begin{pmatrix}
		{1}/{D_u}&         0&             0&       0\\
		0&         {1}/{D_v}&             0&       0\\
		0&                 0&             1&       0\\
		0&                 0&             0&       1
\end{pmatrix}
\cdot
\begin{pmatrix}
		D&  0&  {(N_u-1){\cdot}D_u}/{2}&  0\\
		0&  D&  {(N_v-1){\cdot}D_v}/{2}&  0\\
		0&  0&                          1&  0\\
		0&  0&                          0&  1
\end{pmatrix}
\end{split}\end{equation*}
$M_0$ represents the transformation of the coordinate system from volume to gantry~\cite{wiki:Gantry_(medical)}, $M_{rot}$ denotes the gantry rotation along the Z-axis at the angle ${\beta}$ plus the transpose distance of \emph{d}. 
$M_1$ indicates the projection point on the FPD plane. 
Note that all of the variables are listed in Table~\ref{tbl:cbct-param}.
By expanding Equation~\ref{equ:uv}, the \emph{z} can be written as
\begin{equation}\small \begin{split}
\label{equ:z}
z = d+sin(\beta)(\tilde{i}-{(N_x-1)}/{2}){\cdot}D_x-cos(\beta){\cdot}(\tilde{j}-{(N_y-1)}/{2}){\cdot}D_y
\end{split}\end{equation}
Clearly, the \emph{z} is independent of $\hat{k}$ and equals d+$y_{ab}$, where $y_{ab}$ is the Y coordinate of line $\overline{ab}$, which is parallel to the Z-axis in Figure~\ref{fig:CT.geometry}a.

\subsubsection{\bf{Reducing the Cost of Computing the Projections}}
In this section, we present a method to improve the performance of back-projection by reducing the projections computational cost.
Algorithm~\ref{alg:bp} is adopted in a number of CBCT applications, e.g. open-source libraries (OSCaR~\cite
{rezvani2007ff}, RabbitCT~\cite{rohkohl2009rabbitct}, RTK~\cite{rit2014reconstruction}), and literature~\cite{xu2007real,zhao2009gpu,okitsu2010high,hofmann2014performance}.
Based on \emph{Theorem-2} and \emph{Theorem-3}, the improved back-projection is illustrated in Algorithm~\ref{alg:bp-v1}. The computational cost for computing the projections becomes a factor of 1/6 of the original FDK algorithm (Alg~\ref{alg:bp} line 6), since we only compute half of the {$N_z$} dimension (Algorithm~\ref{alg:bp-v1} line 11), and one of inner products for the two 1$\times$4 vectors (Algorithm~\ref{alg:bp-v1} line 12) instead of the three inner products in the original algorithm (Algorithm~\ref{alg:bp} line 6). Algorithm~\ref{alg:bp-v1} shows the optimized variables highlighted in gray color.
More specifically, the values of \emph{u} and $W_{dis}$ are reused for $N_z$ times, and only half of \emph{y} is computed directly.
Zhao et al.~\cite{zhao2009gpu} discussed a rotational symmetry in the projection layout. We do not adopt this methodology since it is impractical for pipeline processing in terms of the high latency of collecting the four projections. 
\begin{algorithm}[t]
  \footnotesize
  \caption{Proposed Back-projection algorithm. 
  The optimized variables are highlighted in gray color.
  }
  \label{alg:bp-v1}
  \begin{algorithmic}[1]
  	\Require{$P_i, Q_i, N_p, N_x, N_y, N_z, i\in[0,N_p)$}
    \Ensure{$I$}\Comment{reconstructed 3-dimensional volume}
	\State{$\tilde{I}{\gets}0$}\Comment{$\tilde{I}$ initialization}
	\For {$s\in[0,N_p)$}
    	\State{$\bm{\tilde{Q_s}}{\gets}Q_s^T$}\Comment{\textcolor{red}{transpose 2D matrix}}
    	\For {$j\in[0,N_y)$}
        	\For {$i\in[0,N_x)$}
            	\State{$t\gets[i, j, 0, 1]$}              
                \State{\hllightgray{$[x,z]{\gets}[\inner{P_s[0]}{t},\inner{P_s[2]}{t}]$}}\Comment{\textcolor{red}{2 inner product}}
                \State{$f{\gets}1.0/z $}
                \State{{${u}$}$\gets x{\cdot}f $}
                \State{{$W_{dis}$}$\gets f^2$}\Comment{weighting}
                \For {$k\in[0,~$\hllightgray{${{N_z}/{2}}$}~)} \Comment{symmetric geometry}
                    \State{\hllightgray{$y{\gets}\inner{P_s[1]}{[i, j, k, 1]}$}}\Comment{\textcolor{red}{1 inner product}}
                    \State{{${v}$}${\gets}y{\cdot}f $} \Comment{compute v only}               
                    \State{$\tilde{I}(k,j,i) \gets \tilde{I}(k,j,i)+W_{dis}{\cdot}interp2(\tilde{Q_s}, v, u)$}
                    \State{$\tilde{k}{\gets}N_z-1-k$}\Comment{symmetric geometry}
                    \State{{${\tilde{v}}$}${\gets}N_v-1-v$}\Comment{symmetric geometry}
                    \State{$\tilde{I}(\tilde{k},j,i) \gets \tilde{I}(\tilde{k},j,i)+W_{dis}{\cdot}interp2(\tilde{Q_s}, \tilde{v}, u)$}
        		\EndFor
            \EndFor
        \EndFor
    \EndFor
    \State{$I{\gets}reshape(\tilde{I})$}\Comment{\textcolor{red}{reshape means changing data layout}}
  \end{algorithmic}
\end{algorithm}
\subsubsection{\bf{Improving Data Locality: Data Layout \& Loops}}
This section discusses how we leverage the proposed algorithm to derive an implementation that improves the data locality. To increase the cache hits for accessing the volume (\emph{I}) and projection (\emph{Q}) in Algorithm~\ref{alg:bp}, the proposed  Algorithm~\ref{alg:bp-v1} adjusts the memory layout and re-organizes the loops for $N_p$, $N_x$, $N_y$, and $N_z$. The original \emph{I} uses an i-major layout as Figure~\ref{fig:CT.geometry}b shows. The \emph{I} becomes k-major in the proposed algorithm. Based on \emph{Theorem-2}, the proposed memory layout is more cache-friendly for data access, since the data buffers of both the projections and the volume can be accessed contiguously, as shown by the marked variables of $\tilde{Q_s}$ and $\tilde{I}$ in Algorithm~\ref{alg:bp-v1}.

Note that this memory layout is general to all kinds of processors, i.e. CPUs, GPUs, and Xeon Phi. To improve the data locality, the authors in \cite{zinsser2013systematic,lu2016cache} implemented the back-projection kernel on GPUs by organizing the loops as Algorithm~\ref{alg:bp-v1}, such that they compute along the z-axis first. However, that method does not optimize for the layout of arrays \emph{Q} and \emph{I}. Hence, their implementation becomes further complex since one has to rely on CUDA's 2D-texture cache to improve the data locality. It is noteworthy that the time required to transpose a projection (Algorithm~\ref{alg:bp-v1} line 3) is a small fraction of the filtering (or back-projection) stage and thus, we do not discuss its effect on the overall performance of FDK in the later sections.

\lstset{
 	language = C++, breaklines = true, breakindent = 10pt, basicstyle = \ttfamily\scriptsize, commentstyle = {\itshape \color[cmyk]{1,0.4,1,0}}, classoffset = 0, keywordstyle = {\bfseries \color[cmyk]{0,1,0,0}}, stringstyle = {\ttfamily \color[rgb]{0,0,1}}, frame = trbl, framesep=5pt, numbers = left, stepnumber = 1, xrightmargin=7pt, xleftmargin=8pt, numberstyle = \tiny, tabsize = 3, captionpos = t, directivestyle={\color{black}},  emph={int,char,double,float,unsigned}, emphstyle={\color{blue}},
}
\begin{figure}[t!]
\centering
\begin{minipage}[c]{0.49\textwidth}
\begin{lstlisting}[caption = {The proposed back-projection CUDA kernel. Constant memory-optimized ProjMat is defined to store the 3${\times}$4 projection matrixes. The \emph{dot} function computes inner product,
%of size 1${\times}$4, 
\emph{mad} is the fused-multiply-and-addition intrinsic, \emph{interp2} function is implemented as Algorithm~\ref{alg:subpixel}. The batch of projections (defined as $N_{batch}$) is 32.}, label = listing:cuda-shfl-bp]
__constant float4 ProjMat[32][3];//32 3x4 proj-matrix
__global void shflBP(float* vol, int3 vol_size, const float* img, int3 img_dim){
	int laneId = threadIdx.x & 31;
	//(k,j,i) is the position of a voxel 
	int i = blockIdx.x*blockDim.x + threadIdx.x;
	int j = blockIdx.y*blockDim.y + threadIdx.y;
	int k = blockIdx.z*blockDim.z + threadIdx.z;
	float4 vec = make_float4(k, j, i, 1);
	//Z, U are 2 registers
	float Z, U;
	if (laneId < img_dim.z) {
		Z = 1/dot(ProjMat[laneId][2], vec); //inner product
		U = dot(ProjMat[laneId][0], vec)*Z; //inner product
	}
	float sum = 0, _sum = 0, v, _v;
	int _i = vol_dim.z - 1 - i; //geometry symmetric
	for (int s = 0; s<img_dim.z; s++) {
	  //get register value via shuffle 
	  float u = __shfl_sync(0xffffffff, U,  s);
	  float f = __shfl_sync(0xffffffff, Z,  s);
	  float Wdis=f*f;                   //weighting
	   v = dot(ProjMat[s][1], vec)*f;   //inner product
     _v = img_dim.x - 1 - v;           //geometry symmetric
     //sub-pixel interpolation & weighting
	  sum=mad(interp2(img,s,img_dim.x,img_dim.y, v,u),Wdis, sum);
	 _sum=mad(interp2(img,s,img_dim.x,img_dim.y,_v,u),Wdis,_sum);
	}
	//update voxels
	  pVolOut                             [ i] +=  sum;
    (pVolOut + (k*vol_dim.y+j)*vol_dim.z)[_i] += _sum;
}
\end{lstlisting}
\end{minipage}
\end{figure}
\subsection{Back-projection on GPU}\label{our-cuda-implementation}\label{sec:cuda-bp}
In this section, we introduce the proposed back-projection implementation in CUDA and elaborate on the optimization of the CUDA kernel using the shuffle intrinsic.
\subsubsection{\bf{CUDA Implementation}} This section describes the proposed CUDA implementation.
We implement the proposed back-projection kernel (called shflBP in Listing~\ref{listing:cuda-shfl-bp}), which can be used in all generations of Nvidia GPUs from Kepler to Volta architectures. 
In Listing~\ref{listing:cuda-shfl-bp}, the detailed CUDA kernel is presented.
We use global memory (as introduced in Section~\ref{sec:cuda}) to store the 3D volume (see the variable of \emph{vol} in Listing~\ref{listing:cuda-shfl-bp}) due to its huge size. 
{
Though the CUDA unified memory~\cite{cudaToolkit} is also an attractive choice for storing 3D volume, we avoid using it due to its unstable performance, which varies from CUDA version to version as reported in~\cite{awan2018oc,manian2019characterizing}.
}
To further reduce the computation cost by sharing data between threads, we take advantage of the shuffle instruction to perform intra-warp communication as introduced in Section~\ref{sec:cuda}. 
The shflBP kernel processes a batch of projections in one pass. This benefits the overall performance in many aspects:
\begin{enumerate*}[label=(\Roman*)]
\item decreasing the access count of the volume data which is stored in the global memory.
\item increasing in-register accumulation for back-projection computation.
\item eliminating the overhead of launching multiple CUDA kernels.
\end{enumerate*}
Note that this strategy is also applied in the widely used image library RTK~\cite{rit2014reconstruction}.
However, as explained earlier, the proposed algorithm outperforms RTK by reducing the computational cost and improving data locality.

\subsubsection{\bf{Shuffle-based CUDA kernel: shflBP}} In this section, we explain the details of applying shuffle to our CUDA kernel.  
We employ 2 registers (named {\small\emph{Z}}, {\small\emph{U}}) in the shflBP kernel (Listing~\ref{listing:cuda-shfl-bp} line 10) to store the values of \emph{f} and \emph{u} (Algorithm~\ref{alg:bp-v1} line 7$\sim$9). Hence, the value of variable $W_{dis}$ (Algorithm~\ref{alg:bp-v1} line 10) can be easily obtained as \emph{Wdis} in Listing~\ref{listing:cuda-shfl-bp} line 21.
We avoid the use of shared or global memory all together by taking advantage of the shuffle instruction to realize the data communication between threads ( Listing~\ref{listing:cuda-shfl-bp} line 19$\sim$20). Since the scope of the shuffle instructions is limited to a single CUDA Warp, the strategy of sharing data as Algorithm~\ref{alg:bp-v1} is adjusted to a single CUDA Warp. 
To the best of our knowledge, exchanging registers by shuffle is superior in terms of effective throughput, in comparison to using the shared memory~\cite{chen2018efficient}. Additionally, the shflBP kernel does not require thread block barriers, which is often required when using shared memory. 

\begin{figure*}[t]
\centering
\subfloat[iFDK framework. The input is the 2D projections, the output is the generated 3D volume. vol denotes a sub-volume. $C_i$ and $R_i$ are defined in Table~\ref{tbl:mnFDK-param}.] {
 \includegraphics[width=0.51\textwidth]{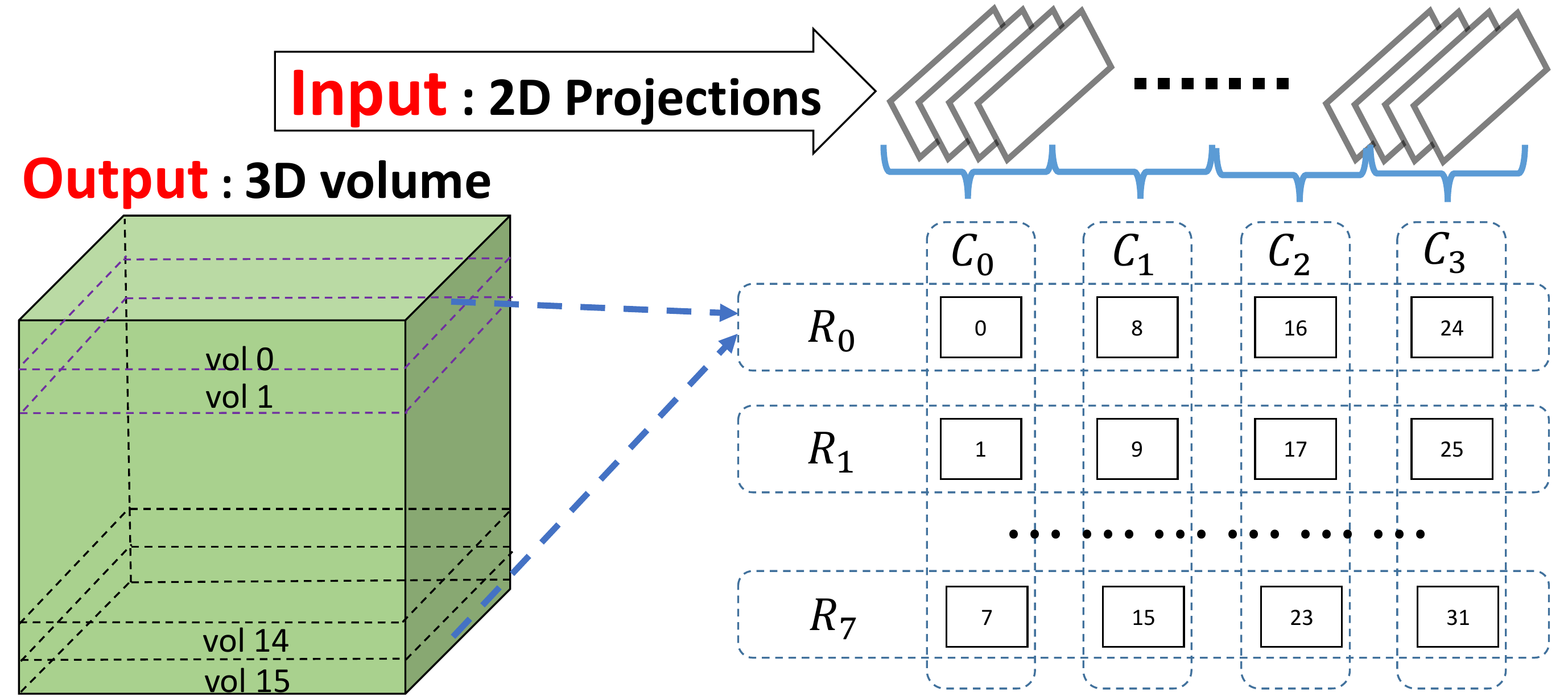}
 \label{fig:2dmpi}
 }
 \hfill\hspace{0.00\textwidth}
 \subfloat[Allgather is performed across the ranks in one column. Reduction is applied only once across the ranks in one row.] {
 \includegraphics[width=0.35\textwidth]{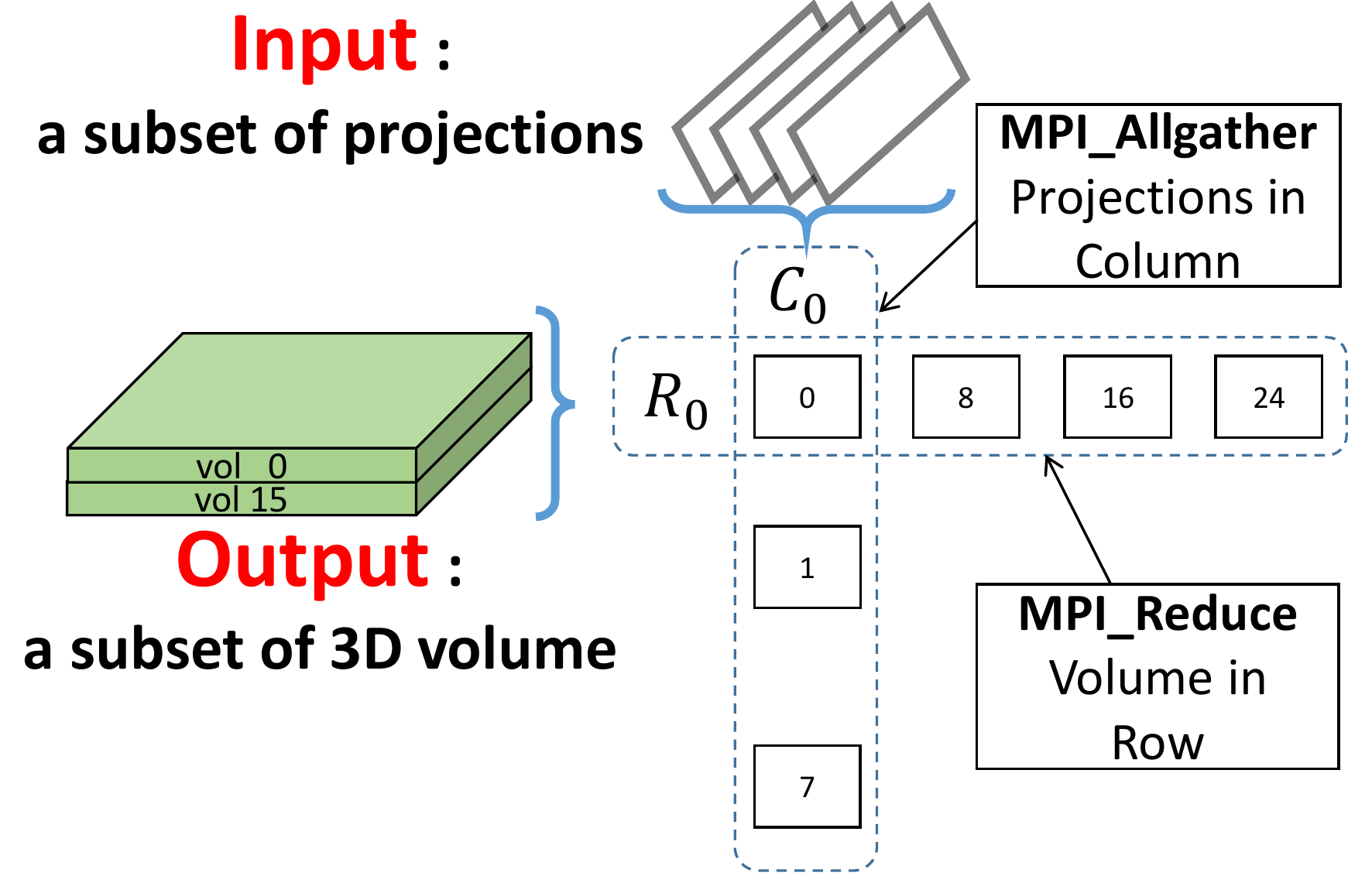}
 \label{fig:sub-2dmpi}
 }
 \hfill\hspace{0.00\textwidth}
\caption{Illustrative example of the iFDK framework. 32 MPI ranks are arranged in a 2D grid of size 4${\times}$8 (R=8, C=4). The input projections are decomposed into 4 groups corresponding to the columns C. The output volume are aggregated from 16 (2*R) sub-volumes.}
\label{fig:mnfdk-framework}
\end{figure*}

\section{Distributed Framework for High-res. Image Reconstruction}\label{sec:mnFDK}
This section presents a distributed framework, namely iFDK, for instant high-resolution image reconstruction. 
The parameters used in iFDK are summarized in Table~\ref{tbl:mnFDK-param}.


\subsection{Design and Implementation of iFDK}\label{sec:mnfdk-imp}
We combine the CPU filtering stage, the GPU-optimized back-projection stage, and MPI as a communication library to scale iFDK to the \O{1000} GPUs. This section elaborates on the design choices and implementation.   

\subsubsection{\bf{2D Grid of MPI Ranks}}
The compute capability and device memory capacity are limited in a single GPU. It is impractical to use a single GPU to generate large volumes (e.g. volumes of size $4096^3$ or $8192^3$). Therefore, we scale the proposed method to take advantage of GPU-accelerated supercomputers to solve those problems.
This section presents the problem decomposition and orchestration of MPI ranks.
To fully utilize the computing resources, i.e. CPUs, GPUs, inter-connectors, we launch multiple MPI ranks within each node (one rank per GPU) to perform the computation and communication concurrently.

 \begin{table}[t]  
      \caption{iFDK parameter list.}        
      \scriptsize
      \centering
      \begin{tabular}{ ?c|l? }
      \tbhline
      \bf{Parameter} & \bf{~~~~~~~~~Descriptions} \\
      \hline
      $R, C$ & the rows and columns of 2D grid of mpi ranks, respectively  \\
      $R_i$ & the $i^{th}$ row of 2D mpi ranks, where $i{\in}[0,R)$ \\
      $C_i$ & the $i^{th}$ column of 2D mpi ranks, where $i{\in}[0,C)$ \\
      $N_{gpu\_per\_node}$ & the number of GPUs per compute node \\
      $N_{cpu\_per\_node}$ & the number of CPUs per compute node \\
      $N_{nodes}$ & the total number of compute nodes \\
      $N_{ranks}$ & the total number of launched ranks \\
      $N_{gpus}$ & the total number of GPUs \\
      $N_{cpus}$ & the total number of CPUs\\      
      $N_{PCIe}$ & the number of PCIe connector per compute node \\
      $N_{proj\_per\_rank}$ & the number of loaded and filtered projections per rank \\
      $N_{cpu\_core}$ & the total number of cores per CPU \\
      $N_{gpu\_mem\_size}$ & the memory capacity of a single GPU \\
      \tbhline
       \end{tabular}  
       \label{tbl:mnFDK-param} 
\end{table}     
The MPI ranks are managed as a 2D-grid of $R$ rows and $C$ columns. The total number of ranks may be expressed as
\begin{equation}\small
\begin{split}
N_{ranks} = C*R
\label{equ:ranks}
\end{split}
\end{equation}
In Figure~\ref{fig:2dmpi}, an example of using iFDK with 32 MPI ranks is presented, where R=8 and C=4.
In iFDK, ranks in each column of the 2D-grid load a subset of projections from the PFS independently. Next, we perform the filtering stage on those projections using the CPUs. The number of projections processed by the ranks in each column of the 2D-grid is $N_p/C$. We use MPI-AllGather to send the filtered projections to neighbor ranks in the same group (namely the same column). 
Hence, the number of projections which are loaded and filtered by a single rank is
\begin{equation}\small
\begin{split} 
N_{proj\_per\_rank} &= {N_p}/{N_{ranks}} = {N_p}/({C*R})
\label{equ:pr}
\end{split}
\end{equation}
Each rank in the same row of the 2D-grid (or the $R_i^{th}$ row) computes the same sub-volumes as Fig~\ref{fig:sub-2dmpi} shows, the final sub-volume that is reconstructed by the $R_i^{th}$ row can be obtained by reducing all of the sub-volumes that are generated by the ranks in the same group (or the $R_i^{th}$ row). 

\subsubsection{\bf{Multi-GPU Management}}
We discuss how the GPUs are managed by MPI ranks in this section. Commonly, in a single compute node, there are multiple GPUs (e.g. ORNL'Summit has six GPUs, LLNL's Sierra and TokyoTech's Tsubame have four GPUs), which are connected to the CPUs by PCIe or NVLink~\cite{li2019evaluating}.
We launch a number of MPI ranks per compute node equivalent to the number of GPUs, i.e. one MPI rank per GPU. The $N_{gpus}$ may be written as
\begin{equation}\small
\begin{split} 
N_{gpus}{\equiv}N_{ranks}
\label{equ:gpus}
\end{split}
\end{equation}
Here, {\scriptsize{$N_{nodes}=N_{ranks}/N_{gpu\_per\_node}$}} is the number of required compute nodes.
Each MPI rank manages a single GPU as follows:
first, we gather the filtered projections that are processed on the CPU, as explained in Section~\ref{sec:filtering-algorithm}. Second, the processed projections are copied from the host to device memory. Third, the back-projection kernel is launched to generate the specified subset of 3D volume. Finally, the computed volume is copied from the device memory to the host. The detailed operations performed inside each rank, by multiple threads, are presented in the next section. 

\subsubsection{\bf{Multiple Threads in a Rank}}\label{sec:multi-threaded-rank}
\begin{figure*}[t]
	\centering
	\subfloat[Three threads execute in parallel in a pipeline fashion and exchange data via two queue-buffers. MPI-AllGather is performed the Main thread.] {
		\includegraphics[width=0.485\textwidth]{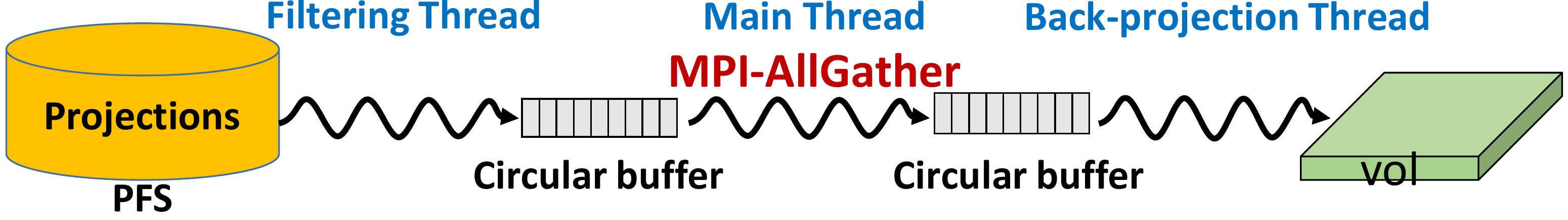}
		\label{fig:a-rank-a}
	}
	\hfill\hspace{0.05\textwidth}
		\subfloat[Reduce all the sub-volumes generated in the previous step to a single final volume to be stored in the PFS by the main thread.] 
	{
		\includegraphics[width=0.40\textwidth]{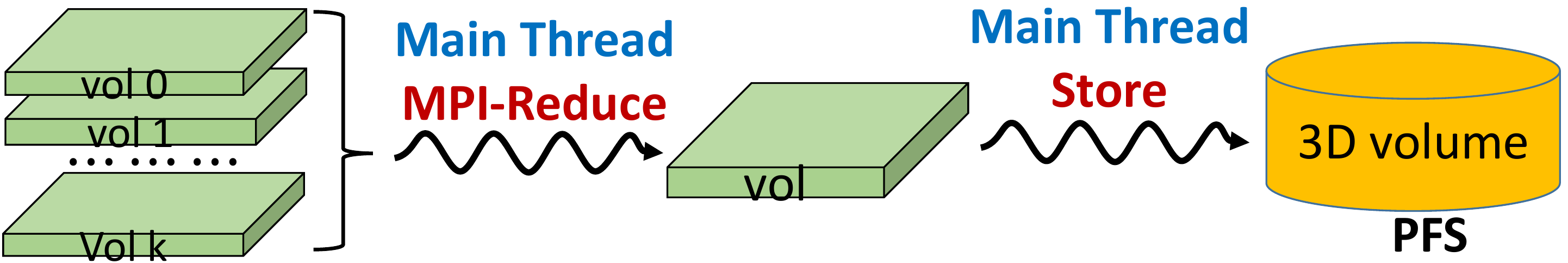}
		\label{fig:a-rank-b}
	}
	
	\subfloat[Example of pipeline to solve $2048^2{\times}4096{\rightarrow}4096^3$ problem using 128 V100 GPUs. R=32 and C=4. Measured execution time for each task is displayed. The Filtering thread processes 32 projections. The Main thread sends 32 projections and gathers 1024 projections using MPI-AllGather. The Bp-thread processes 1024 projections.] {
	\includegraphics[width=1.0\textwidth]{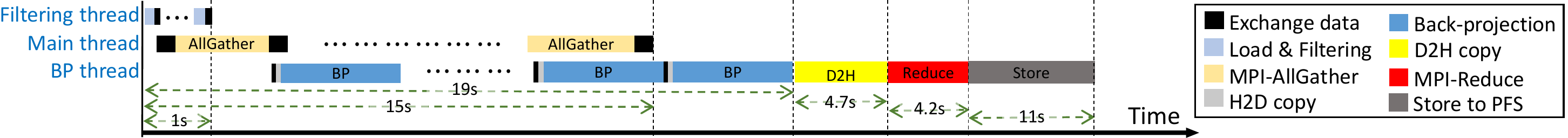}
	\label{fig:overlapping}
	}	
	\caption{Orchestration and Overlapping in iFDK.}
	\label{fig:pipeline}
\end{figure*}

This section presents how multiple threads are orchestrated in each MPI rank. Each rank in iFDK processes several tasks in parallel, e.g. loading projections from PFS, filtering the projections, collective communication, back-projection, and storing the volume to PFS. As Figure~\ref{fig:pipeline} shows, we use three threads to execute those tasks, namely \emph{Main-thread}, \emph{Filtering-thread}, and \emph{Bp-thread} (Back-projection thread) . 
Those threads are created using the \emph{pthread} library, they execute independently and exchange data with each other using circular buffers~\cite{wiki:Circular_buffer}.
The Filtering-thread launches OpenMP threads of number {\scriptsize{(${N_{cpu\_core}*N_{cpu\_per\_node}}/{N_{ranks}}-1$})} to load projections and execute the filtering in parallel.
For each projection, the load and filtering operations are executed within the same OpenMP thread in the sequence that enables immediate processing.

As Figure~\ref{fig:sub-2dmpi} shows, we use the MPI-AllGather collective to gather the specified subset of filtered projections in the Main-thread. At the same time, those filtered projections are dispatched to the designated GPU by Bp-thread for the back-projection computation. Note that for each MPI rank, $N_{proj\_per\_rank}$ times of AllGather operations are required since we process one projection at a time by AllGather.
When all of the filtered projections are finally processed by the GPUs, the generated sub-volume on the GPU device memory is copied back to the host. 
Next, using the MPI-Reduce collective as in Figure~\ref{fig:a-rank-b}, we do a single reduction step to generate the final resulting volume in host memory using the Main-thread (a real example can be found in Figure~\ref{fig:mnfdk-reduce}). 
Finally, the 3D volume is stored in the PFS using multi-ranks with multi-threads.
Note that the volume of size {\small$N_x{\times}N_y{\times}N_z$} is stored as slices of number {\small$N_z$}, the size of each slice is {\small$N_x{\times}N_y$}. There is room for improvement by tuning the size of each slice to optimize for the throughput of storing to the PFS (i.e. tune slice size to optimize for file striping).


\subsubsection{\bf{Orchestration and Overlapping}}\label{sec:insight-overlap}
In order to achieve the optimal overlap, we use three threads to pipeline the computation as Figure~\ref{fig:a-rank-a} shows. Figure~\ref{fig:overlapping} presents a real example of solving a 4K problem using 128 V100 GPUs. To further give insight into the effect of pipelining the computation,
the breakdown of the overlapped computation (namely $T_{compute}$, as defined Section~\ref{sec:prerf-model}) is listed in Table~\ref{tbl:compute}. The value of $\delta$>1 indicates that we achieve the goal of improving the overall performance by overlapping different stages, i.e. filtering, collective communication, and back-projection.
This overlapping scheme improves overall performance since back projection is the main bottleneck, and hence we get a streaming benefit from the overlapping. On the other hand, overlapping the tasks after the back-projection (i.e. the device to host copy, reduction, and storing to PFS) does not guarantee any performance improvement (for the price of complexity introduced). Nonetheless, overlapping after the back-projection remains to be one of the points of investigation in future work.

\subsubsection{
\bf{Configuration of Parameter R}}\label{sec:R}
This section discusses how to select the optimal value for the parameter R: the number of rows in the 2D mesh of MPI ranks. Based on the design of iFDK (in Figure~\ref{fig:2dmpi}), R may be expressed as
\begin{equation}\small \begin{split}
R = sizeof(float)*N_x*N_y*N_z/N_{sub\_vol}
\label{equ:R}
\end{split} \end{equation}
where $N_{sub\_vol}$ is the size of sub-volume. For a specified number of GPUs (or ranks), we minimize the value of R and maximize the value of C for three reasons:
{
\begin{enumerate*}[label=(\Roman*)]
    \item We can efficiently use the limited device memory since each GPU would be able to compute a volume of size {\scriptsize$sizeof(float)*N_x*N_y*N_z/R$};
    \item We can achieve higher computational performance by generating larger volumes. As Table~\ref{tbl:bp-eva} shows, regarding the specified input, bigger output (smaller $\alpha$) results in better performance for the back-projection kernel;
    \item To the specified workload of {\scriptsize $N_u{\times}N_v{\times}N_p{\rightarrow}N_x{\times}N_y{\times}N_z$}, there is a linear relationship between the runtime and $N_p$. Since we decompose the workload to sub-tasks ({\scriptsize $N_u{\times}N_v{\times}\frac{N_p}{C}{\rightarrow}N_x{\times}N_y{\times}N_z$}) of number C, it is essential to maximize the value of C to decrease the runtime of each sub-task.
\end{enumerate*}
}

In addition, the value of R is often power of two and is constrained by the memory capacity of a GPU as follows
\begin{equation*}\small \begin{split}
sizeof(float)*(\frac{N_x*N_y*N_z}{R}+N_u*N_v*N_{batch}){\leq}N_{gpu\_mem\_size}
\label{equ:mem-capacity} \end{split} \end{equation*}
where $N_{batch}$=32 as Listing~\ref{alg:bp-v1} shows. 
Given $N_{gpu\_mem\_size}$=16GB in the GPU generation we use, for the high-resolution reconstruction problems we target, $N_{sub\_vol}$=8GB is adopted.

\subsection{Performance Model}\label{sec:prerf-model}
We discuss a performance model intended to analyze the impact of different parameters on performance and predict the potential peak performance.
\subsubsection{\bf{Micro-benchmarks.}}
We use micro-benchmarks to measure peak throughput parameters of constant values in our model (for a given system).  
$BW_{load}$ and $BW_{store}$ are the aggregate throughput of reading and writing to the PFS, respectively. Both of them are measured by LLNL IOR~\cite{shan2007using}. $TH_{flt}$ is the throughput of filtering computation that we measure by running the filtering kernel on the target CPU.
$TH_{bp}$ is the throughput of our back-projection kernel as measured on the target GPU (we also report it in Table~\ref{tbl:bp-eva} to give perspective to readers interested in only the back projection kernel). 
$TH_{trans}$ is the throughput of transposing a volume on GPU.
The $TH_{AllGather}$ and $TH_{Reduce}$ are the throughput of MPI-AllGather and Reduce APIs, respectively. Both of them are measured by Intel mpi-benchmarks.
$BW_{PCIe}$ is the throughput of data transfer between the host and device memory via a single PCI-e and is measured by Nvidia's tool called bandwidthTest.

\subsubsection{\bf{iFDK Performance Model}}\label{sec:model-peak-performance}
Given execution time required for: reading projections from storage $T_{load}$, filtering projections $T_{flt}$, communicating all projections by MPI-AllGather $T_{AllGather}$, copying filtered projections from host to device $T_{H2D}$, back-projection $T_{bp}$, transposing the sub-volume $T_{trans}$, moving the sub-volume from device memory to host $T_{D2H}$, reducing the sub-volume $T_{Reduce}$, and storing volume to PFS $T_{store}$. 
Those variables can be written as
\begin{equation}\small \begin{split}
&T_{load} = {sizeof(float)*N_u*N_v*N_p}/{BW_{load}}
\label{equ:bw-load}
\end{split} \end{equation}
\begin{equation}\small \begin{split}
T_{flt} = \frac{N_p}{N_{nodes}*TH_{flt}} = \frac{N_p*N_{gpu\_per\_node}}{C*R*TH_{flt}}
\label{equ:bw-flt}
\end{split} \end{equation}
\begin{equation}\small\textbf{} \begin{split}
&T_{AllGather} = {N_p}/({C*R*TH_{AllGather}})
\label{equ:bw-allgather} 
\end{split} \end{equation}
\begin{equation}\small \begin{split}
&T_{H2D} = \frac{sizeof(float)*N_{gpu\_per\_node}*N_u*N_v*N_p}{C*BW_{PCIe}*N_{PCIe}}
\label{equ:bw-bp} 
\end{split} \end{equation}
\begin{equation}\small \begin{split}
&T_{bp} = T_{H2D}+{N_p}/({C*TH_{bp}})
\label{equ:bw-bp} 
\end{split} \end{equation} 
\begin{equation}\small \begin{split}
T_{trans} = {sizeof(float)*N_x*N_y*N_z}/({R*TH_{trans}}) 
\label{equ:bw-trans}
\end{split} \end{equation}
\begin{equation}\small \begin{split}
T_{D2H} = \frac{sizeof(float)*N_{gpu\_per\_node}*N_x*N_y*N_z}{R*BW_{PCIe}*N_{PCIe}} 
\label{equ:bw-d2h} 
\end{split} \end{equation}
\begin{equation}\small \begin{split}
T_{reduce} = {sizeof(float)*N_x*N_y*N_z}/({R*TH_{Reduce}}) 
\label{equ:bw-reduce} 
\end{split} \end{equation}
\begin{equation}\small \begin{split}
T_{store} = {sizeof(float)*N_x*N_y*N_z}/({BW_{store}}) 
\label{equ:bw-reduce} 
\end{split} \end{equation} 
As Figure~\ref{fig:pipeline} shows, the three threads compute in parallel such that most of the computation and data movement is overlapped. Additionally, the filtering thread launches multiple OpenMP threads to perform the loading and filtering operations concurrently as Section~\ref{sec:mnfdk-imp} explained. The filtering operation can be perfectly overlapped since {\scriptsize$T_{load}+T_{flt}{\ll}T_{bp}$}, as the example shows in Figure~\ref{fig:overlapping}.
Here, the execution time required by the three threads may be expressed as
\begin{equation}\small \begin{split} 
T_{compute} = \max(T_{load}, T_{flt}, T_{AllGather}, T_{bp})
\label{equ:compute} 
\end{split} \end{equation}
The time required to transpose, copy and reduce the volume is
\begin{equation}\small \begin{split} 
T_{post} &= T_{trans}+T_{D2H}+T_{reduce}+T_{store}\\
         &~{\approx}~T_{D2H}+T_{reduce}+T_{store}
\label{equ:post} 
\end{split} \end{equation}
Note that the $T_{trans}$ is a small value ({\scriptsize{$T_{trans}{\ll}{T_{D2H}/{10}}$}} is observed) that could be ignored . The total execution time is
\begin{equation}\small \begin{split}
T_{runtime} &= T_{compute}+T_{post}\\
            &~{\approx}~T_{compute}+T_{D2H}+T_{reduce}+T_{store}
\label{equ:runtime} \end{split} \end{equation}

\subsubsection{\bf{Conclusions from the Performance Model}}

{
We conclude the discussion of our performance model as follows.
\setlength{\leftmargini}{15 pt}
\begin{enumerate*}[label=(\Roman*)]
\item {\bf{Scalability:}} The performance of iFDK scales with the number of GPUs ($N_{gpus}$) since $T_{flt}$, $T_{AllGather}$ and $T_{bp}$ are inversely proportional to C (as in Equation~\ref{equ:bw-flt}$\sim$\ref{equ:bw-bp}), C is proportional to $N_{gpus}$ (as in Equation~\ref{equ:ranks} and Equation~\ref{equ:gpus}), where R is a value that is minimized to meet the constraints  described in Section~\ref{sec:R}, where $T_{post}$ is a constant value in $T_{runtime}$.
\item {\bf{Potential peak performance:}} According to Equation~\ref{equ:runtime}, the potential peak performance of the computation ($T_{runtime}$) can be used to quantify the efficiency of our implementation. 
\end{enumerate*}
}

\section{Evaluation}\label{sec:Evaluation} 
This section lists the experimental environment, reports the performance of the proposed algorithms and discusses the scalability of the framework. 
{
\subsection{How Performance Was Measured}
This section lists the experimental environment and discusses how the performance was measured. 

\paragraph{\bf{HPC system and environment}}
AIST's ABCI~\footnote{System is ranked $8^{th}$ on the TOP500 list as of June 2019.} supercomputer
is used for our evaluation. 
Each compute node is equipped with two Intel Xeon Gold 6148 CPUs (20 Cores), 384GB memory, four Tesla V100 GPUs (16GB RAM) through PCIe gen3$\times$16 and two InfiniBand EDR HCAs.
ABCI uses CentOS 7.4 for the operating system and mounts a 6.6PB GPFS shared storage. The iFDK framework is implemented by Nvidia CUDA 9.0 (CUDA driver version: 410.104), Intel Performance Libraries 2018.2.199 (includes MPI and IPP). The compilers nvcc-9.0 and mpicc (included in MPI library) are used to compile the CUDA kernel and host code, respectively. The compiler option (-gencode arch=compute\_70,code=sm\_70) is applied in nvcc-9.0 for Nvidia's Volta architecture. 

\paragraph{\bf{Measurement methodology}}
Since the performance of image reconstruction is independent of the content of projections or volume, we apply the standard Shepp-Logan phantom~\cite{shepp1974fourier} to generate a variety of projections by the forward-projection tool in RTK library. An example that is reconstructed by our framework can be found in Fig.~\ref{fig:mnfdk-reduce}. 
We use single precision for all projections, volumes, and runs. 
For output verification, we use the image processing tool ImageJ~\cite{abramoff2004image} to render the generated 3D volumes, then inspecte them manually. We also use profiled runs to investigate the density value of each voxel. Finally, we compare the output with the volumes generated by the RTK library on the CPU, the Root Mean Square Error (RMSE) is less than 10e-5.
We use the OS-independent function \emph{cudaEvent} to measure the execution time of CUDA kernels and employ \emph{MPI\_Wtime} to measure the application's runtime. Each of the reported results is averaged by 100 runs. Finally, we report the performance in the unit of GUPS, as defined in Section~\ref{terminology}.
}

\subsection{Performance of the Back-projection Kernel}\label{sec:eva-fdk}


This section reports the performance of the proposed back-projection kernel on Tesla V100 GPU by comparisons with a collection of kernels (listed in Table~\ref{tbl:bp-kernels}).
The latest RTK 1.4.0 implementation (called RTK-32 in Table~\ref{tbl:bp-eva}) is used with 32-bit precision (versus the default 8-bit precision). 
{Note that our target is to instantly generate high-resolution volumes. We demonstrate that we could achieve this goal while using high image quality: \emph{we do not sacrifice the quality by using lower precision.}} 

The kernel function kernel\_fdk\_3Dgrid of RTK is strictly implemented as defined in Algorithm~\ref{alg:bp-v1}. The original RTK limits the maximum count of projections to 16, we extend it to 32. Also, we adjust its interpolation function (Algorithm~\ref{alg:subpixel}) to the precision of 32-bits that uses 2D-Layered cache without linear interpolation: namely using the cudaFilterModePoint parameter for the texture function. Regarding L1 cache-optimized access for projections in Table~\ref{tbl:bp-kernels}, the \_\_ldg intrinsic is applied. 

\begin{table}[t]  
      \caption{Back-projection kernel characteristics. "Texture cache" and "L1 cache" mean accessing projections via 2D-Layered texture and L1 cache, respectively. "Transpose projection" and "Transpose volume" mean transposing the projections and volume, respectively. RTK-32 is the imrpoved RTK kernel. The other kernels are shflBP kernel (as in Listing~\ref{listing:cuda-shfl-bp}) with different characteristics.}        
      \scriptsize
      \begin{tabular}{ ?c|cccc?}
      \tbhline
                       & \bf{Texture cache} & \bf{L1 cache}    & \bf{Transpose projection}    & \bf{Transpose Volume}      \\
      \hline
      {RTK-32}       & $\checkmark$      &     \xmark            &     \xmark     & \xmark       \\       
      {Bp-Tex}       & $\checkmark$      &     \xmark            &     \xmark     & $\checkmark$       \\         
      {Tex-Tran}     & $\checkmark$      &     \xmark            &     $\checkmark$      & $\checkmark$       \\          
      {Bp-L1}        & \xmark            &     \xmark            &     $\checkmark$      & $\checkmark$        \\         
      {L1-Tran}      & \xmark            &     $\checkmark$      &     $\checkmark$      & $\checkmark$       \\               
      \tbhline
       \end{tabular}  
       \label{tbl:bp-kernels} 
\end{table} 

\begin{table}[t]
      \caption{Back-projection kernel performance on Tesla V100 GPU.
      1k and 2k mean 1024 and 2048, respectively.
      $\alpha$ is defined as the ratio of input to output problem size.
      The characteristics of evaluated CUDA kernels are listed in Table~\ref{tbl:bp-kernels}. 
      } 
      \centering 
      \scriptsize 
      \begin{tabular}{?r|r|rrrrr?}
      \tbhline
     \bf{FDK poblems} & \multirow{2}{*}{\bf{$\alpha$}} &\bf{RTK-32} & \bf{Bp-Tex} & \bf{Tex-Tran} & \bf{Bp-L1} & \bf{L1-Tran} \\
      {$(pixel{\rightarrow}voxel)$} & &{(GUPS)} & {(GUPS)} & {(GUPS)} & {(GUPS)} & {(GUPS)} \\ 
      \tbhline
      $512^2{\times}1k{\rightarrow}128^3$  & 128 &65.3 & 38.8 & 46.5 & 23.7 & \bf{118.0} \\ 
      $512^2{\times}1k{\rightarrow}256^3$  &  16 &107.4	& 96.2	& 98.9	& 28.0	& \bf{188.6} \\ 
      $512^2{\times}1k{\rightarrow}512^3$  &   2 &115.1	& 105.8	& 106.1	& 34.0	& \bf{206.0} \\ 
      $512^2{\times}1k{\rightarrow}(1k)^3$   &   1 &118.1	& 107.3	& 107.3 & 64.9	& \bf{211.4} \\ 
      $512^2{\times}1k{\rightarrow}(1k)^2{\times}2k$ & 1/8 &N/A & 107.4 & 107.6	& 112.1	& \bf{212.7} \\ \hline
      $(1k)^3{\rightarrow}128^3$ & 512 &41.9 & 13.8 & 13.5 & 5.7 & \bf{27.2} \\ 
      $(1k)^3{\rightarrow}256^3$ & 64 &77.4 & 35.9 & 43.2 & 12.8 & \bf{83.7} \\ 
      $(1k)^3{\rightarrow}512^3$ &  8 & 115.7 & 95.5 & 98.1 & 25.1 & \bf{190.3} \\ 
      $(1k)^3{\rightarrow}(1k)^3$ &   1 &117.9 & 105.8 & 105.8 & 34.0 & \bf{205.7} \\ 
      $(1k)^3{\rightarrow}(1k)^2{\times}2k$ &  1/2&N/A & 106.3 & 106.5 & 65.0 & \bf{207.9} \\ \hline 
      $(2k)^2{\times}1k{\rightarrow}128^3$ & 1024 &16.1 & 5.8 & 8.5 & 2.8 & \bf{7.7} \\ 
      $(2k)^2{\times}1k{\rightarrow}256^3$ & 256 &38.6 & 12.7 & 12.6 & 4.4 & \bf{24.1} \\ 
      $(2k)^2{\times}1k{\rightarrow}512^3$ & 32 &80.2 & 35.5 & 42.5 & 13.9 & \bf{81.6} \\ 
      $(2k)^2{\times}1k{\rightarrow}(1k)^3$ & 4 &116.9 & 94.4 & 97.8 & 23.9 & \bf{186.9} \\ 
      $(2k)^2{\times}1k{\rightarrow}(1k)^2{\times}2k$ &  1&N/A & 102.9 & 104.1 & 33.4 & \bf{198.7} \\ 
      \tbhline
      \end{tabular}  
      \label{tbl:bp-eva} 
\end{table}

\begin{figure*}[t]
\centering
 \subfloat[Strong scaling for $2048^2{\times}4096{\rightarrow}4096^3$. R=32, C=${N_{gpus}}/{32}$.] {
 \includegraphics[width=0.485\textwidth]{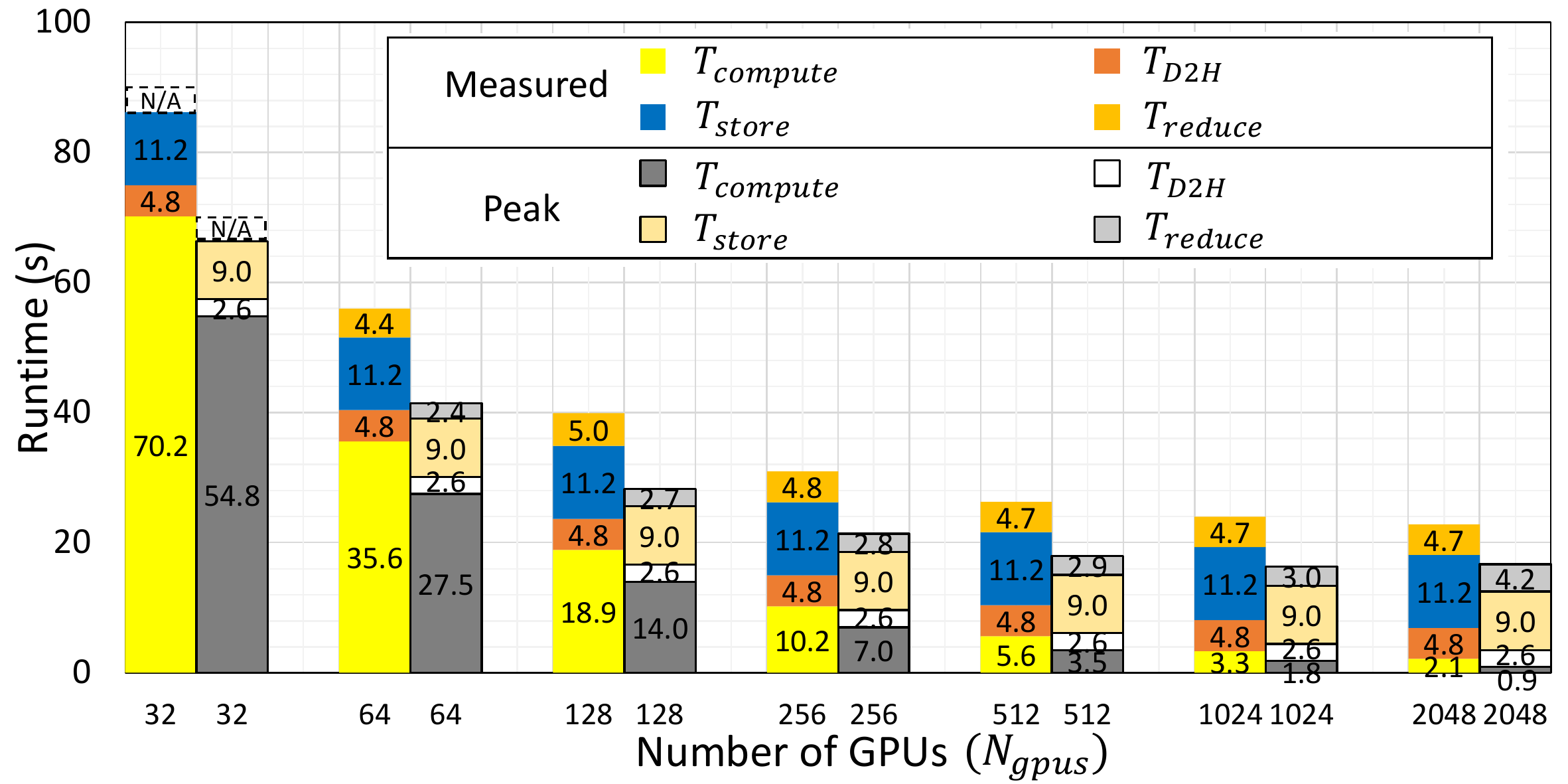}
 \label{fig:strong4k-stacked}
 }
 \hspace{0.001\textwidth}
  \subfloat[Strong scaling for $2048^2{\times}4096{\rightarrow}8192^3$. R=256, C=${N_{gpus}}/{256}$.] {
 \includegraphics[width=0.485\textwidth]{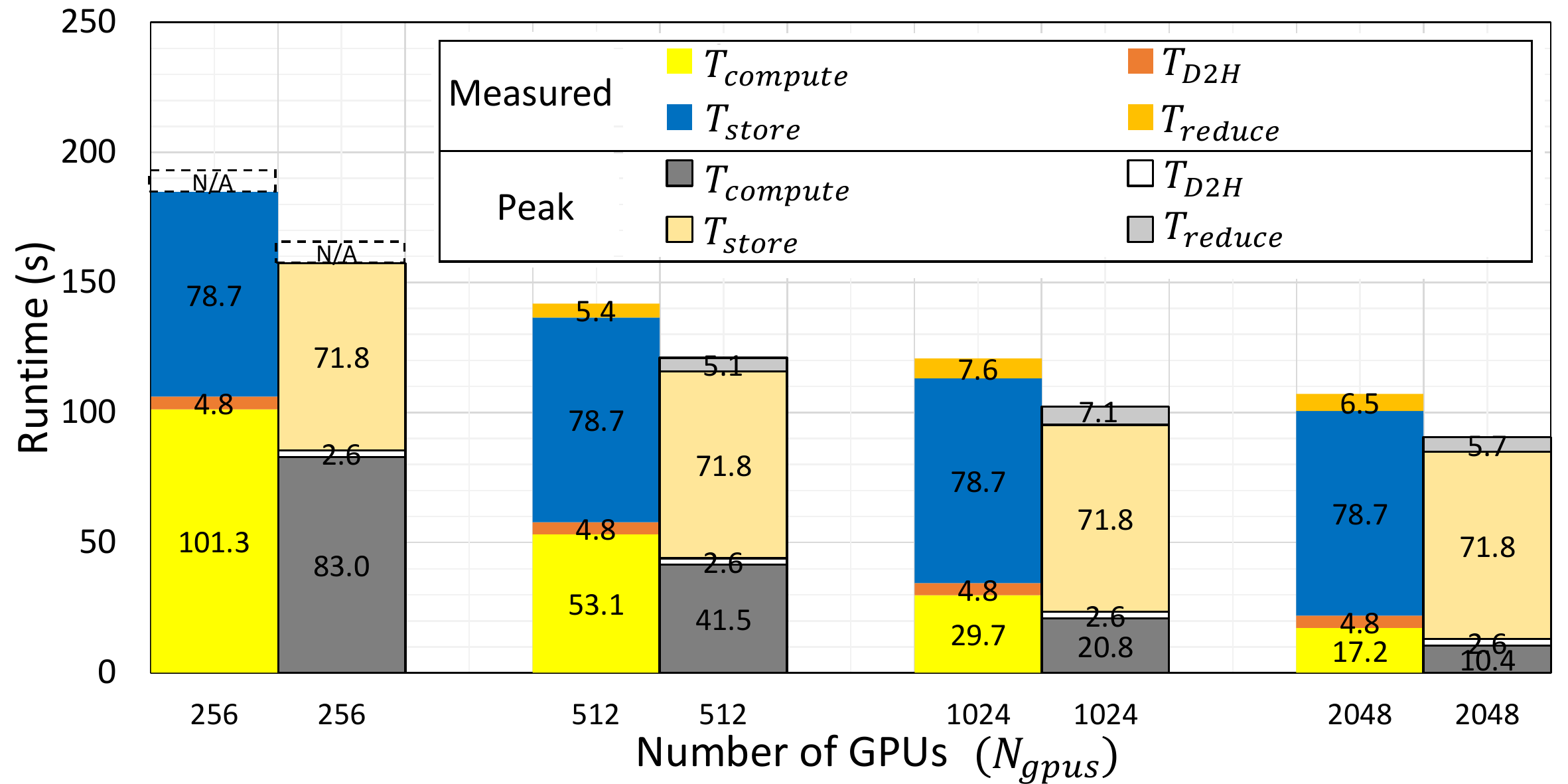}
 \label{fig:strong8k-stacked}
 }
 
 \subfloat[Weak scaling for $2048^2{\times}N_p{\rightarrow}4096^3$. $N_p$=16*$N_{gpus}$, R=32, C=$N_{gpus}$/32.] {
 \includegraphics[width=0.485\textwidth]{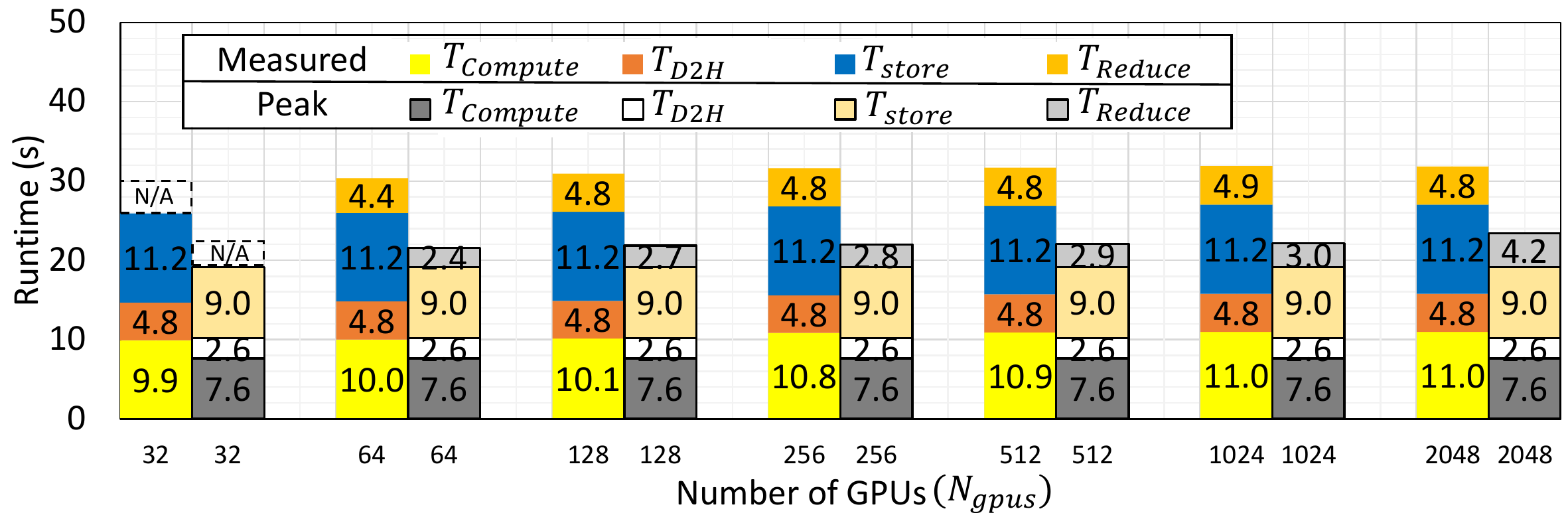}
 \label{fig:weak4k-stacked}
 }
 \hspace{0.001\textwidth}
  \subfloat[Weak scaling for $2048^2{\times}N_p{\rightarrow}8192^3$. $N_p$=4*$N_{gpus}$, R=256, C=$N_{gpus}$/256.] {
 \includegraphics[width=0.485\textwidth]{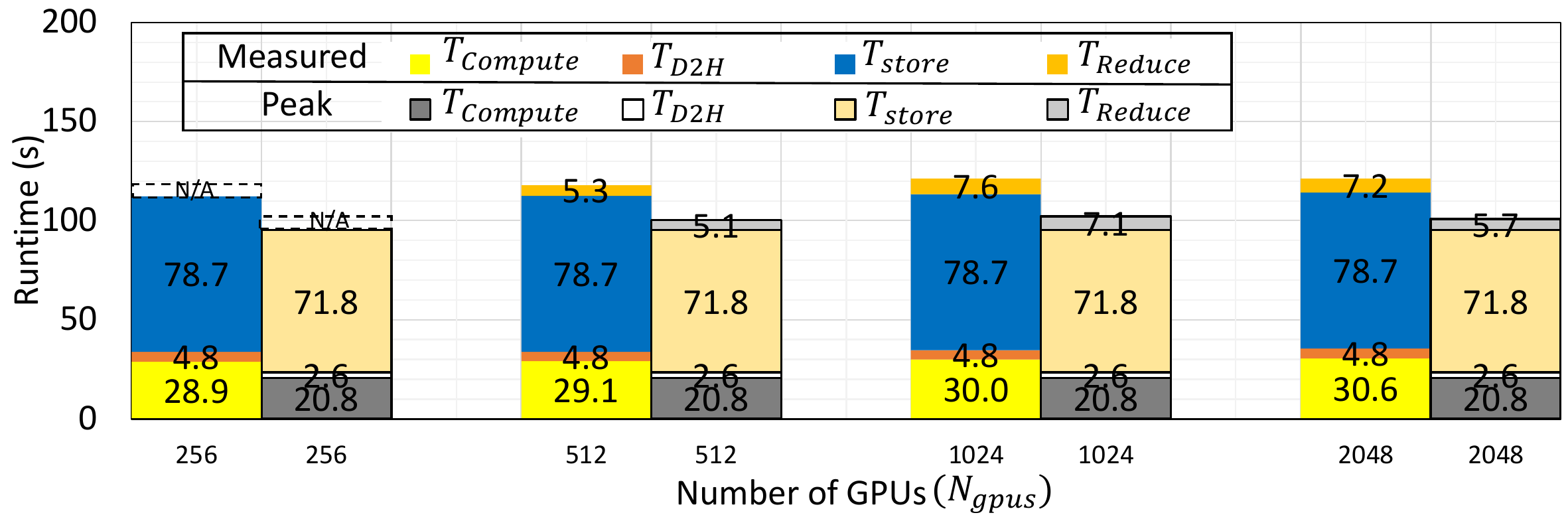}
 \label{fig:weak8k-stacked}
 }
\caption{Scaling iFDK. 
"Reduce" denotes volume reduction. "store" indicates storing the volume to PFS. "D2H" means GPU$\rightarrow$CPU copy. Loading projections from PFS, AllGather communication are overlapped with $T_{compute}$. CPU$\rightarrow$GPU copy is included with $T_{compute}$. $T_{reduce}$ is N/A when C=1 (no inter-rank reduction occurs). The high performance of the proposed back-projection kernel exposes other bottlenecks (e.g. I/O).}
\label{fig:scalability}
\end{figure*}
\begin{figure}[t]
  \begin{center}
    \includegraphics[clip,width=0.48\textwidth]{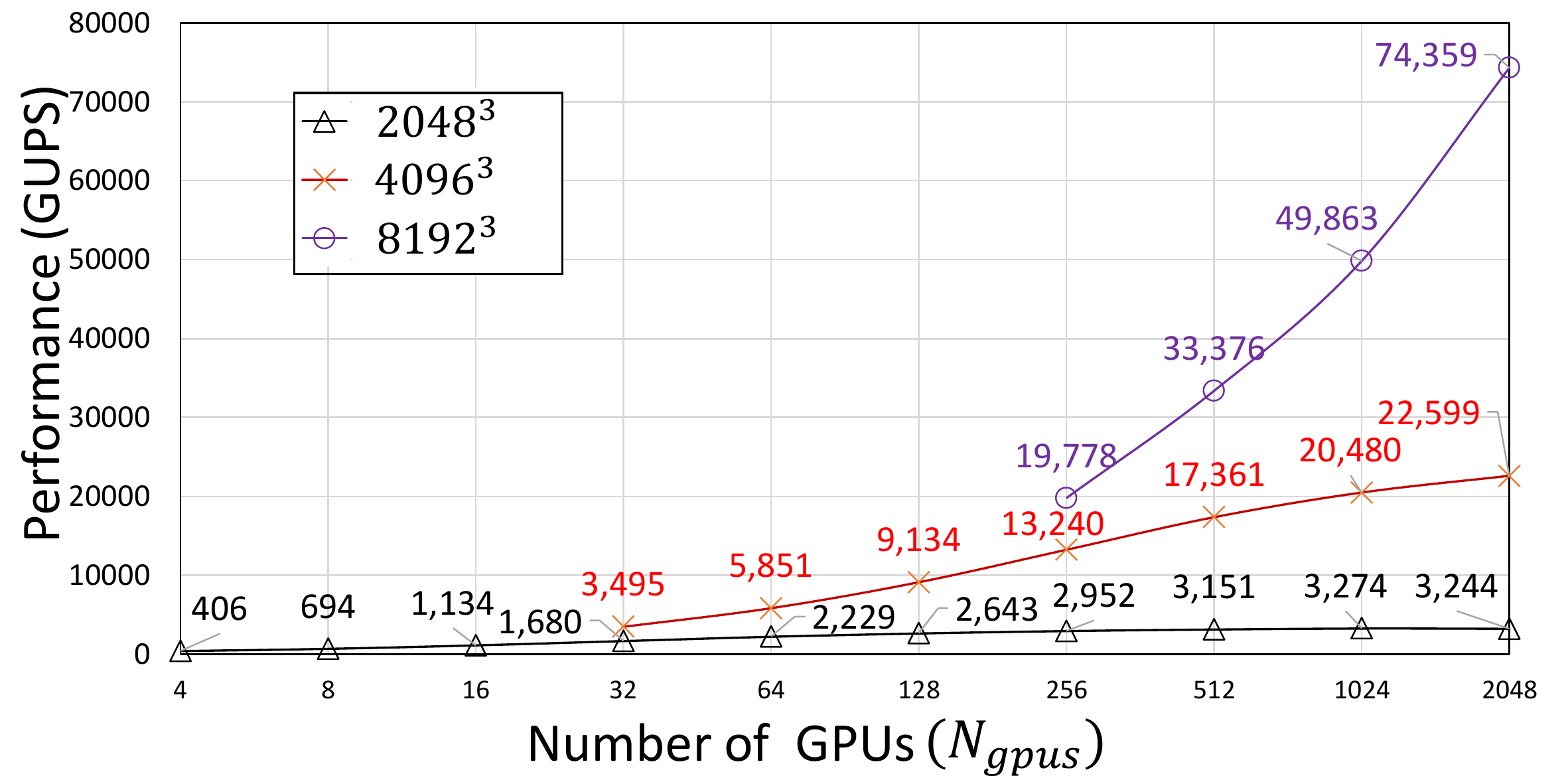}
    \caption{Performance in a unit of GUPS for problem with input size $2048^2{\times}4096$ and three different output sizes $2048^3$, $4096^3$ and $8192^3$. 
    }
    \label{fig:strong-GUPS}
  \end{center}
\end{figure}
The characteristics of the proposed kernels are listed in Table~\ref{tbl:bp-kernels}.
A variety of image reconstruction problems are evaluated on those kernels, the performance in GUPS is listed in Table~\ref{tbl:bp-eva}. 
The time required to move data between host and device is not included in the execution time for the calculation of GUPS.
Note that in most applications, the value of $\alpha$ is typically very small, often less than 1. As seen in Table~\ref{tbl:bp-eva}, the proposed CUDA kernel, namely "L1-Trans", outperforms the other kernels. The performance advantage is due to the improved data locality and efficient intra-warp communication.
As Table~\ref{tbl:bp-eva} shows, the size of the output of RTK cannot be bigger than 8GB since RTK employs a dual buffer technique to store the volume while the maximum memory capacity of Tesla V100 is 16GB in our testbed. By inspecting the performance in Table~\ref{tbl:bp-eva} we can observe the following:
{
\begin{enumerate*}[label=(\Roman*)]
\item When comparing the performance difference between Bp-Tex and Tex-Trans, it appears that the transpose operation of the projection has a minor effect on the hit-rate of the 2D-Layered texture cache.
\item However, the transpose operation noticeably contributes to the Bp-L1 cache hit-rate as observed by comparing the performance of Bp-L1 and L1-Trans.
\item The proposed kernel outperforms the most commonly used production library, for high-resolution image reconstruction.
\end{enumerate*}
}


\subsection{Scaling and Performance}\label{sec:strong-scalability}
This section presents the performance and scalability of iFDK. 
In the scaling experiments, each GPU computes a sub-volume of 8GB (namely $N_{sub\_vol}$). According to Equation~\ref{equ:R}, R=32 and R=256 are used for generating volumes of $4096^3$ and $8192^3$, respectively. In the special case of iFDK that C=1,  $T_{reduce}$ becomes zero since the volume reduction is not required (shown as "N/A" in each sub-figure of Figure~\ref{fig:scalability}).

{
Note that we focus on discussing the impact of the volume parameters $N_x$, $N_y$, $N_z$ and number of projections $N_p$ on the performance and scalability in the following sections. The reason is that we cannot target weak scaling by increasing the input image sizes $N_u$ or $N_v$.
As Algorithm~\ref{alg:filter} shows, only the filtering computation is dependent on $N_u$ and $N_v$. However, the back-projection, which is the main kernel that we scale, is independent of $N_u$ or $N_v$ (see Algorithm~\ref{alg:bp}). This becomes clear from the performance metrics definition of GUPS as in Section~\ref{terminology}, i.e. no dependency on $N_u$ or $N_v$.

}

\subsubsection{\bf{Strong Scaling}}
In Figure~\ref{fig:strong4k-stacked} and Figure~\ref{fig:strong8k-stacked}, the stacked execution time of $T_{compute}$, $T_{post}$ (namely $T_{reduce}$+$T_{D2H}$) and $T_{store}$ are displayed. Note that all $T_{load}$, $T_{flt}$ and $T_{AllGather}$ are included in $T_{compute}$ as Equation~\ref{equ:compute} shows. Since the value of R is fixed, the value of C in iFDK increases in proportion to $N_{gpus}$. This results in $T_{compute}$ decreasing inversely in proportion to $N_{gpus}$. 
To give further insight of the computational behaviour, 
a breakdown of $T_{compute}$ is listed in Table~\ref{tbl:compute}. As the figure demonstrates, iFDK follows the same scaling behavior of the potential peak performance. 

\subsubsection{\bf{Weak Scaling}}
Figure~\ref{fig:weak4k-stacked} and Figure~\ref{fig:weak8k-stacked} show the weak scaling. The evaluated number of GPUs (namely $N_{gpus}$) is up to 2,048. 
In both figures, each rank loads and processes 16 and 4 projections, respectively.
We use an MPI-AllGather operation to get filtered projections from the ranks in the same column group (as in Figure~\ref{fig:sub-2dmpi}). In the back-projection stage, each rank processes 128 and 1024 projections, respectively. 

\begin{figure}[t]
  \begin{center}
    \includegraphics[clip,width=0.48\textwidth]{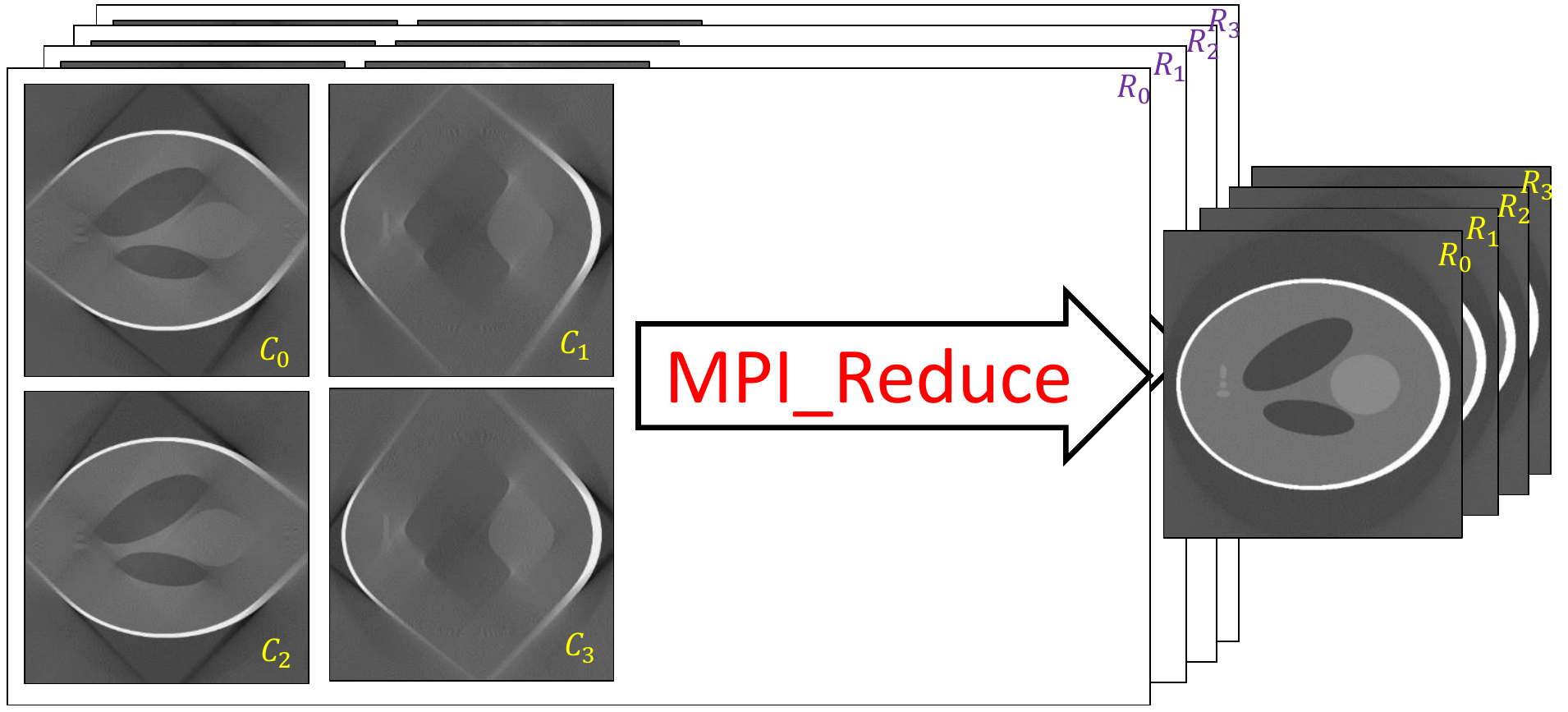}
    \caption{Results of volume reduction in an experiment done in iFDK.  
    The problem is $2048^2{\times}4096{\rightarrow}2048^3$.
    iFDK parameters are R=4 and C=4. 
    $N_{gpus}$ is 16, 
    performance is 1,134 GUPS.
    }
    \label{fig:mnfdk-reduce}
  \end{center}
\end{figure}
\subsubsection{\bf{Performance}}
The potential peak and achieved performances are displayed in Figure~\ref{fig:scalability}. Note that $T_{trans}$ is as small as 0.29s and thus, it is included in $T_{D2H}$ for simplifying the stacked bar figures. 
The peak performance is projected by our performance model. 
For example, the overall potential peak performance (namely $T_{runtime}$) is obtained by the benchmark values as follows.
The peak bandwidth of a single PCIe$\times$16 (namely $BW_{PCIe}$) is 11.9GB/s, the projected time required to copy data of size 32GB (8G*4) from device memory to the host by dual PCI-e connectors is $\approx$2.6s (namely $T_{D2H}$). The projected time required to reduce 8GB of data by dual InfiniBand per node is $\approx$2.7s (namely $T_{reduce}$). 
The peak sequential write bandwidth of GPFS (namely $BW_{store}$) is 28.5GB/s and thus, the projected time required to store data of size 256GB and 2TB is $\approx$ 9.0s and 87.7s (namely $T_{store}$), respectively.

On average, we can achieve {76\%} of the potential peak performance. Upon analyzing the performance gap, we found the following. For $T_{compute}$, the data exchange between the three threads orchestrating the workflow, as Figure~\ref{fig:pipeline} shows, can have some overhead that contributes to the gap. The overhead of intra-thread data movement within the back-projection thread also has an overhead: before copying data from host to GPU memory, it is necessary to gather at least 32 projections as a batch. In addition, memory management, building the circular buffers (as in Figure~\ref{fig:a-rank-a}) also contribute to the gap in a minor way. For $T_{reduce}$, we confirmed that the gap is due to MPI-Reduce being called only once: the first call to the collective is typically slower, which is why benchmarks, like the one we use, pre-run few iterations before measurements. For $T_{D2H}$, the architecture of the compute node can cause contention on the PCIe switch feeding two GPUs (two PCIe switches for four GPUs). For $T_{store}$, it appears from our investigation that the minor gap is the effect of volume slices written to PFS not tuned to the ideal stripe size.
Figure~\ref{fig:strong-GUPS} shows the overall performance of iFDK in GUPS. We use the entire end-to-end execution time as defined in Equation~\ref{equ:runtime}.
Figure~\ref{fig:mnfdk-reduce} shows an example that uses MPI-Reduce primitive to generate a volume of $2048^3$.
iFDK scales better in generating the volume of $8192^3$ than $4096^3$. It is due to better device utilization in the former. This becomes clear from observing the relationship between the coefficient $\alpha$ and performance, in Table~\ref{tbl:bp-eva}. Note that the $2048^2{\times}4096{\rightarrow}4096^3$ and $2048^2{\times}4096{\rightarrow}8192^3$ problems can be instantly solved within {\fourKperf} seconds and {\eightKperf} minutes, respectively. Finally, we report that the performance of the back-projection kernel on a single GPU is $\approx$200 GUPS (without using mixed precision), as shown in Table~\ref{tbl:bp-kernels}. This performance of the back-projection kernel is the main reason a time-to-solution in \O{10} seconds for high-resolution problems on \O{1000} GPUs becomes possible. It is noteworthy that all computation, running on both CPUs and GPUs, use single precision to maintain high fidelity solutions.


\begin{table}[t]  
      \caption{Details of $T_{compute}$ in Fig.~\ref{fig:strong4k-stacked} and Fig.~\ref{fig:strong8k-stacked}. $T_{load}$ is included in $T_{flt}$. $\delta$ is defined as {\tiny{$\delta=\frac{T_{flt}+T_{AllGather}+T_{bp}}{T_{compute}}$}}. 
      }  
      \centering
      \tiny
      \begin{tabular}{ ?c|cc|ccc|c|c? }
      \tbhline 
      \bf{volume} & \multirow{2}{*}{\bf{$N_{gpus}$}}&  \multirow{2}{*}{ \bf{$N_{cpus}$}} &\bf{$T_{flt}$}& \bf{$T_{AllGather}$}& \bf{$T_{bp}$} & \bf{$T_{compute}$} & \multirow{2}{*}{{\bf{$\delta$}}} \\
      (voxel)     &                 &                  &(s)                   & (s)           &(s)                 &(s)          &                                  \\
      \tbhline
      \hline
\multirow{4}{*}{$4096^3$} & 32  & 16   & 1.4   &   31.4  & 54.8  & 70.2  & \bf{1.2}\\
                          & 64  & 32   & 0.8   &   20.7  & 27.5  & 35.6  & \bf{1.4}\\      
                          & 128 & 64   & <0.7  &   15.2  & 14.0  & 18.9  & \bf{1.6}\\
                          & 256 & 128  & <0.7  &    7.4  &  7.0  & 10.2  & \bf{1.5}\\      
      \hline
\multirow{4}{*}{$8192^3$} & 256  & 128  & <0.7  &  46.9 &  83.0  & 101.3  & \bf{1.3}\\   
                          & 512  & 256  & <0.7  &  26.9  & 41.5  & 53.1  & \bf{1.3}\\         
                          & 1024 & 512  & <0.7  &  17.0  & 20.8  & 29.7  & \bf{1.3}\\    
                          & 2048 & 1024 & <0.7  &   8.6  &  10.4  & 17.2  & \bf{1.2}\\                    
      \tbhline
       \end{tabular}  
       \label{tbl:compute} 
\end{table} 


\subsubsection{\bf{Scaling to 8K FDK problems}}
The iFDK is general to any general sizes of FDK problems, due to the two-dimensional problem decomposition methodology. 
{
At present, the common perspective in CT imaging is that processing $8192^3$ volumes is 
highly demanded but not feasible (as discussed extensively by Martz et al.~\cite{X-ray:imaging})}. Nonetheless, we conducted experiments with iFDK on 8K volumes. Using iFDK, with 2,048 GPUs, the $2048^2{\times}4096{\rightarrow}8192^3$ problem is solved within {\eightKperf} minutes. This time-to-solution is inclusive of I/O: it includes storing the volume of size \emph{2TB} to PFS in 79s. 
{
Note that it is rare to find high precision scanners for capturing images with the quality necessary for 8K resolution, and hence it is rare to find a CT system generating the 3D volume of 8K. 
}
Yet we conducted 8K experiments for the sake of evaluating the scalability of iFDK and demonstrating that 8K image reconstruction is technically attainable.


\subsubsection{\bf{Performance Model Accuracy}}
Table~\ref{tbl:compute} lists the detailed execution time of each component in the framework. Our analysis of the results and the resulting observations are as follows:
{
\vspace{-3pt}
\setlength{\leftmargini}{15 pt}
\begin{enumerate}[label=(\roman*)]
\item {\small$\delta>1$}. This indicates that the orchestration and pipelining methodology as introduced in Section~\ref{sec:insight-overlap} is effective in improving the overall performance.
\item {\small$T_{AllGather}{<}T_{bp}$}. MPI-AllGather is employed to share the filtered projections within the $C_i$ group as in Figure~\ref{fig:2dmpi}, $T_{AllGather}$ can be considered as the overhead of filtering computation. Fortunately, it can be overlapped with the back-projection stage as Figure~\ref{fig:pipeline} shows. Furthermore, this justifies the strategy of minimizing R (as discussed in Section~\ref{sec:R}) to generate as large as possible sub-volumes on the GPUs.
\end{enumerate}
}
{
\section{Discussion}
In this section, we discuss the impact of iFDK on several real-world applications and show the platforms available for iFDK.
\subsection{Relevance of iFDK to Real-world Applications}\label{sec:disc}
In one of the main reference textbooks of image reconstruction~\cite{X-ray:imaging} (Section 10.7.1), Martz et al. address in detail the necessity of high-resolution image reconstruction, and give an open question about challenge of high-resolution image reconstruction (we quote): {\bf{"What happens if we start manipulating $(6k)^3$ and $(8k)^3$ volumes?"}}. The work in this paper tackles this challenge by paving the way for generating high-resolution volumes instantly using algorithmic innovation and HPC best practices.

High-resolution image reconstruction is essential since it can expose more detailed information (i.e. geometry, intensity) and provide benefits to the throughput of the industrial inspection and scientific research. The remainder of this section gives concrete examples for the state-of-art innovation in commercial products for high-resolution image reconstruction.

In~\cite{GOM-CT}, the industrial CT (named GOM CT) is equipped with FPD of resolution $3008\times2512$ and is available to reconstruct 3D volumes larger than $3000^2\times5000$ (note the specification: the voxel is 2um$\sim$80um and the measuring area is 240mm$\times$240mm$\times$400mm).
In~\cite{xth450}, the CT system (named XTH450) is equipped with FPD of variant resolutions (i.e. $2000^2$, $4000^2$) and is built for turbine blade and casting inspection.
In~\cite{shimadzu:4k}, the CT system (named inspeXio SMX-225CT) is equipped with FPD of resolution ($4096^2$) and is used for defect inspection.

Scientific research is also witnessing a boom of interest in high-resolution image reconstruction. For instance, Bice et al.~\cite{bicer2015rapid} proposed rapid tomographic image reconstruction by large scale parallelization (up to 32k cores) to meet the critical demands from scientists.  since they require \emph{quasi-instant} feedback in their experiments for rapidly checking results and adjusting experimental configurations while using CT images. This instant capability is critically demanded in scanning objects with huge volumes (e.g. motor engine, human brain~\cite{nowogrodzki2018world}) and complex details (e.g. the body of insects).

In the industrial field, the CT is widely used for defect inspection and reverse engineering~\cite{hirakimoto2002microfocus,fang2013application}. 
Asadizanjani et al.~\cite{asadizanjani2015non} employed micro CT for non-destructive PCB reverse engineering.
The importance of high-resolution in reverse engineering is critical for huge objects with complex structures.



}

\subsection{Platforms for iFDK}
This section discusses the potential for the practical use case of iFDK in many fields, i.e. medical, industrial, and scientific. 
The proposed back-projection algorithm and CUDA implementation can be applied in a number of iterative solvers (i.e. ART, MLEM, MBIR), which are popular methodologies in medical imaging for low dose image reconstruction. In addition, it can provide benefits for real-time CT systems, e.g. 4D-CT~\cite{keall2004acquiring}. 

\subsubsection{{
AWS HPC}}
The methods proposed by iFDK are not limited in use to top-tier HPC systems accessible only to a limited number of researchers. A simple calculation shows that
generating a 4K volume as in Figure~\ref{fig:strong4k-stacked} can be done, for example, on Amazon's AWS HPC offerings for the cost of less than \$100. That is when using 256 \emph{p3.8xlarge} EC2 instances (with four V100 GPUs per node, similar to the system used in this paper). This uses on-demand pricing with billing timed by seconds at the price of \$12.24 per hour (March 2019 US east Ohio region). This accounts for the low-performance network in AWS (10Gbps) by assuming factors of a slow down over the reported performance. 
Note that we do not consider moving the volumes out of the cloud since the volume visualization and data analysis can be performed in the same cloud platform.
We acknowledge that policy and privacy issues would complicate the use of public clouds in medical and industrial CT. Yet we emphasize that, from the technical perspective, the methods used in iFDK make the practical and affordable instant high-resolution image reconstruction feasible.  

{
\subsubsection{Nvidia DGX-2}
In order to avoid the privacy issues mentioned earlier, one could use an on-premise dense GPU box to achieve the performance reported by iFDK, for reasonably high resolution. For instance, the Nvidia DGX-2~\cite{dgx2} is an appropriate alternative considering it is equipped with 16 Tesla V100 GPUs, the total GPU memory of 512GB, system memory of 1.5TB, and internal storage of 30TB. Those specifications would allow iFDK, for instance, to tackle 4K problems within a minute (projected by the results shown in Figure~\ref{fig:strong4k-stacked}) without privacy concern. In addition, the DGX-2 has the fast NVSwitch~\cite{li2019evaluating} interconnect between GPUs, and a high capacity SSD: iFDK would even perform better due to improvements in the communication and I/O. Finally, it is important to mention that the price of DGX-2 is relatively low when considering the prices of high-end CT instruments.}

\section{Related Work}\label{sec:related-works}
Researchers have been working for a long time to optimize CT image reconstruction algorithms.
Wu et al. presented an Application Specific Integrated Circuits (ASIC) solution, which is efficient in computation but expensive in development~\cite{wu1991asic}. 
The works in~\cite{coric2002parallel,xue2006acceleration,subramanian2009c,henry2012fpga}
applied Field-Programmable Gate Array (FPGA)
to accelerate the image reconstruction computation. Programming FPGA by HDL language
is complex and thus, the trend in recent years is to use high-level synthesis approaches, e.g. OpenCL~\cite{siegl2011opencl,held2016analysis,martelli20173d}. 
Both Treibig et al. and Hofmann et al. optimized FDK by SIMD instruction set extensions
and achieved record performance for CPUs~\cite{treibig2013pushing, hofmann2014comparing}.
The performance engineering for \emph{RabbitCT}~\cite{rohkohl2009rabbitct} on the Intel Xeon Phi accelerator by Johannes et al. demonstrated a promising approach for optimizing back-projection algorithms~\cite{hofmann2014performance}. 
Mostly, the prior work focused on parallelizing image reconstruction algorithms on the specified accelerators.

The prior work in~\cite{zhao2009gpu,lu2016cache} focused on decomposing the output problems into several sub-volumes in order to solve the problem out-of-core. 
Wang et al.~\cite{wang2017massively} used a 69,632-core distributed system to accelerate the iterative reconstruction and achieved 1,665$\times$ speed-up over the baseline. The authors in~\cite{blas2014surfing} use up to 6 GPU to reconstruct the volume of 2K and 4K via the FDK method. Lack of details, e.g. computational precision and I/O, complicates direct comparisons with the work mentioned above. 
An iterative image reconstruction algorithm called Distributed MLEM~\cite{cui2013distributed} scaled up to 16 GPUs on multi-GPU clusters.
Hidayetoglu et. al~\cite{8425161} present a massively-parallel solver (iterative method) for tomographic image reconstruction, their solution scales up to 4,096 GPUs. 


\section{Conclusion} \label{sec:Conclusion}
The filtered back-projection method is indispensable in {most of the} practical CT systems. In this work, we propose a novel and general FDK algorithm that reduces the computational cost and improves data locality.
{Using CUDA, we implement an efficient back-projection kernel.}
We further propose a distributed framework that leverages the heterogeneity in modern systems to solve high-resolution image reconstruction problems (e.g. 4K, 8K) in tens of seconds.
{
More specifically, we optimize the filtering computation and back-projection on CPUs and GPUs, respectively. 
We perform the inter-node hierarchical computation by MPI collective communication primitive. 
{
The experimental results demonstrate the performance and scalability of iFDK, and also validate our performance model.
}
For future work, we intend to investigate compression and visualization of the high-resolution volumes. 
We also plan to provide a real-time image reconstruction cloud service. 
}



\section*{Acknowledgment}
This work was partially supported by JST-CREST under Grant Number JPMJCR1687. 
Computational resource of AI Bridging Cloud Infrastructure (ABCI) provided by National Institute of Advanced Industrial Science and Technology (AIST) was used. 
A part of the numerical calculations were carried out on the TSUBAME3.0 supercomputer at Tokyo Institute of Technology.
We would like to thank Endo Lab at Tokyo Institute of Technology for providing a Tesla V100 GPU Server.
\bibliographystyle{ACM-Reference-Format}
\bibliography{bibliography}

\end{document}